\documentclass[Afour,sageh,times]{sagejarxiv}

\usepackage{moreverb,url}

\usepackage[colorlinks,bookmarksopen,bookmarksnumbered,citecolor=red,urlcolor=red]{hyperref}
\usepackage{csquotes}
\usepackage{natbib}
\usepackage{listings}
\newcommand\BibTeX{{\rmfamily B\kern-.05em \textsc{i\kern-.025em b}\kern-.08em
		T\kern-.1667em\lower.7ex\hbox{E}\kern-.125emX}}

\def\volumeyear{}

\newcommand{\rem}[1]{}
\usepackage{todonotes}

\definecolor{darkblue}{rgb}{0.0, 0.0, 1}
\definecolor{darkgreen}{rgb}{0.0, 0.4, 0}

\newcommand{\revision}[1]{{\color{black}{#1}}}
\newcommand{\revisionHS}[1]{{\color{black}{#1}}}

\usepackage{etoolbox}
 
\usepackage[TS1,T1]{fontenc}

\begin{document}

\runninghead{}

\title{Biased processing and opinion polarization: experimental refinement of argument communication theory in the context of the energy debate}

\author{Sven Banisch\affilnum{1}\affilnum{2}\affilnum{*} and Hawal Shamon\affilnum{3}}
\affiliation{
\affilnum{1} Max Planck Institute for Mathematics in the Sciences, Leipzig, Germany\\
\affilnum{2} Institute of Technology Futures, Karlsruhe Institute of Technology, Germany\\
\affilnum{3} Institute of Energy and Climate Research - Systems Analysis and Technology Evaluation (IEK-STE), Forschungszentrum J\"{u}lich, Germany\\
\affilnum{*} Corresponding author (sven.banisch@universecity.de)}



\begin{abstract}
\revisionHS{In sociological research, the study of macro processes, such as opinion polarization, faces a fundamental problem, the so-called micro-macro problem. To overcome this problem, w}e combine empirical experimental research on biased argument processing with a computational theory of group deliberation in order to clarify the role of biased processing in debates around energy. The experiment reveals a strong tendency to consider arguments aligned with the current attitude more persuasive and to downgrade those speaking against it. This is integrated into the framework of argument communication theory in which agents exchange arguments about a certain topic and adapt opinions accordingly. We derive a mathematical model that allows to relate the strength of biased processing to expected attitude changes given the specific experimental conditions and find a clear signature of moderate biased processing. We further show that this model fits significantly better to the experimentally observed attitude changes than the neutral argument processing assumption made in previous models. Our approach provides new insight into the relationship between biased processing and opinion polarization. At the individual level our analysis reveals a sharp qualitative transition from attitude moderation to polarization. At the collective level we find (i.) that weak biased processing significantly accelerates group decision processes whereas (ii.) strong biased processing leads to a persistent conflictual state of subgroup polarization. While this shows that biased processing alone is sufficient for \revision{the emergence of} polarization, we also demonstrate that homophily 
may lead to intra-group conflict at significantly lower rates of biased processing.
\end{abstract}

\maketitle


\section{Introduction}

Social processes can currently be observed around the world in which controversies over various issues are coming to a head. For example, while some members of society are strongly in favor of a political decision-maker, others are strongly opposed to the same political leader (e.g. Trump, Erdogan, Putin, Lukashenko). The same processes can be identified worldwide for other objects of attitude, such as migration movements, measures to contain the COVID pandemic or climate change and its cause(s). These developments hold potential for danger, as they threaten international but also intranational social cohesion. It is therefore all the more important to understand the mechanisms of such processes in detail. 

\revisionHS{The examination of processes of opinion polarization at the macro level of social aggregates\footnote{In accordance with \citep{Nisbet1970social}, the term social aggregate is defined by \citep[p. 5]{Jasso1980new} as "any physical or conceptual aggregate of humans who are mutually aware of each other and of the aggregate they form."} is challenging from a methodological point of view since they reflect a result of a plethora of individual processes of attitude change at the micro level. Standardized surveys (here also referring to survey experiments) represent a well-established data collection method in social science that allow researchers to measure concepts among analytical units, such as individuals, at particular snapshots in time. However, the knowledge on individuals' characteristics (e.g., attitudes towards coal power plants or wind power stations) is not enough by itself to anticipate phenomena on the social aggregate's macro level that rather occur due to (repeated) interactions between individuals. \cite{Wiley1988micro} has drawn sociologists' attention to this so-called micro-macro problem and stressed out the importance of considering human interactions in social theories \citep[see also][]{Alexander1987micro}. Interactions can give rise to macro-level phenomena, which are greater than the sum of their parts, so-called emergent phenomena.\footnote{Traffic jams are an everyday example of an emergent phenomenon since they cannot be directly anticipated from aggregating the mobility behavior of individuals only, but rather occur by the complex interplay between autonomously acting individuals.} 


Agent based models (ABMs) are a versatile method that allow researchers to overcome the micro-macro-problem, since (repeated) interactions between heterogeneous agents are one, not to say the, core element of this method. In ABMs, researchers define the characteristics of several artificial agents (representing for example individuals) with respect to their behavior and interaction rules as well as the structure of their artificial environment according to their research question. Since agents can also be specified to take their environment's macro-level information into consideration, ABMs are perfectly compatible with Coleman's \revision{macro-micro-macro} research paradigm \citep{Coleman1990foundations}. 
A weak point of ABMs, however, is the choice of an appropriate initial distribution of the agents' properties (e.g., how many oppose coal power stations and how many favor them) as well as of the parameters that govern their behavior and interaction (e.g., how strongly agents are considered to favor interaction with like-minded others) since those may heavily determine the macro level outcome and process of a model \citep{Maes2017random}. This drawback can be overcome by deriving agents' characteristic distributions as well as the parameter space of behavior and interaction rules from empirical research. Against this background, standardized surveys and ABMs are rather complementary than mutually exclusive research methods whose combination is promising to overcome the micro-macro problem in the examination of macro level phenomena in Sociology \citep[cf. also][]{Shamon2018surveys}.}



A variety of theoretical models have been developed to understand the mechanism behind the emergence of consensus, polarization and conflict over opinion. Theoretical approaches such as social influence network theory \citep{Friedkin1999choice,Friedkin2011social} and social feedback theory \citep{Banisch2019opinion, Gaisbauer2020dynamics} put a primary focus on how the structure of social networks impacts the dynamical evolution of attitudes in a group or a population. It is well known from this research that network segregation and community structure favor diversity and polarization. Other computational studies explain (sub)group polarization based on the homophily principle \citep{Lazarsfeld1954friendship,McPherson2001birds} by which the propensity of social exchange depends on the similarity of opinions. These models show that preferences for interaction with similar others may lead to persistent plurality \citep{Carley1991theory,Axelrod1997dissemination,Hegselmann2002opinion,Banisch2010acs} and polarization \citep{Maes2013differentiation,Banisch2021argument}. 
\revision{Also assumptions about negative social influence by which opinions that are already far from one another will be driven farther apart in interaction have been included and may account for polarization dynamics \citep{Macy2003polarization,Mark2003culture,Baldassarri2007dynamics,Flache2011small}.

All these approaches have in common that they focus in one way or another on social influence processes mediated through a network of social relations\revisionHS{, that is on inter-personal mechanisms}. In this paper we draw attention to an intra-personal process -- namely, biased processing -- and show that neither structural faultlines nor homophily or negative influence are necessary for collective polarization. The intra-individual tendency of biased processing alone is sufficient.

In order to better understand how biased processing contributes to opinion polarization, we combine a survey experiment on argument persuasion with an ABM of group deliberation. Starting from basic assumptions made in the ABM, we derive a simple cognitive model of opinion revision which is then calibrated with data from the survey experiment. This, in turn, enables an empirically informed refinement of the micro assumptions on which the ABM is drawing.
We then characterize the macro-level behavior of the empirically refined ABM with a series of systematic computational experiments.
This provides a detailed qualitative picture of the macro-level effects of biased information processing at the level of individuals.}

Our theoretical model is based on argument communication theory (ACT) advanced by \citep{Maes2013differentiation}. The main idea is that an opinion is a multi-level construct comprised of an attitude layer and an underlying set of arguments \citep[cf.][]{Banisch2021argument}. In repeated interaction agents exchange pro and con arguments about an attitude object and adjust their attitudes accordingly. If this process of argument exchange is coupled with homophily at the level of attitudes this gives rise to the formation of two increasingly antagonistic groups \revisionHS{at the macro-level} which rely on more and more separated argument pools \citep{Sunstein2002law}. As a consequence group opinions become more and more concentrated at the extremes of the opinion scale. ACT has proven very useful to understand the impact of opinion diversity and demographic faultlines in group deliberation processes \citep{Maes2013short,Feliciani2020persuasion} and is also capable to explain how opinions on multiple interrelated topics may align along ideological lines \citep{Banisch2021argument}. The main contribution of this paper is to propose and experimentally validate a refined mechanism of argument exchange that incorporates biased information processing and to show that the group-level predictions of ACT are fundamentally affected when these refined micro-assumptions are incorporated.


Biased argument processing -- also labeled as biased assimilation \citep{Lord1979biased,Corner2012uncertainty,Kobayashi2016relational}, defensive processing \citep{Wood1995working}, refutational processing \citep{Liu2016incompatible} or attitude congruence bias \citep{Taber2009motivated} in the literature -- refers to a person's tendency to inflate the quality of arguments that align with his or her existing attitude on an attitude object whereas the quality of those arguments that speak against a person‘s prevailing attitude are downgraded. A number of empirical studies 
\citep[cf. e.g.][]{Biek1996working,Teel2006evidence,Corner2012uncertainty,Kobayashi2016relational,Shamon2019changing} across different topics and samples have shown that biased processing is a robust cognitive mechanism whenever persons are exposed to a set of opposing arguments on attitude objects. 
In order to integrate this intra-personal tendency of attitude-dependent argument processing, we rely on an empirical study in the context of climate change, and electricity production in particular \citep{Shamon2019changing}. 
In this experiment, attitudes towards six different energy technologies (coal power stations, wind turbines, etc.) \revisionHS{were} measured before and after subjects \revisionHS{had been} exposed to a balanced set of 7 pro and 7 con arguments. Subjects \revisionHS{were} asked to rate the persuasiveness of arguments and their judgments reveal a systematic bias towards attitude-coherent arguments. 

Our cognitive model assumes that this biased evaluation of arguments affects the probability with which arguments are taken up by an agent to a certain degree $\beta$. For a scenario which mimics the experimental design of \cite{Shamon2019changing} as closely as possible, we can derive a statistical model of the expected effects of the argument treatment in which the free model parameters have a clear meaning in terms of mechanisms. We show that this cognitive model fits significantly better to the experimentally observed attitude changes than the neutral argument processing assumption made in all previous ACT models. 

This close alignment of a computational model of information processing and an experiment on argument persuasion sheds light on the relation between biased processing and \revision{opinion polarization in a variety of ways.  At the individual level, it provides a clear understanding of attitude changes in so-called balanced argument treatments where subjects are exposed to a balanced mix of pro and con arguments.
Two forces counteract one another in such a balanced information setting: (i.) there is moderating effect of being exposed to both sides of the opinion spectrum, and (ii.) there is a polarizing effect of filtering arguments in favor of existing beliefs.
}
Empirical studies \citep[e.g.][]{Lord1979biased,Taber2006motivated,Taber2009motivated,Druckman2011framing,Corner2012uncertainty,Teel2006evidence,Shamon2019changing} repeatedly examined whether or not biased processing of balanced arguments \revision{may lead to more extreme attitudes and contribute to polarization tendencies. 
}
Empirical evidence is mixed: while some studies find support for attitude polarization as a consequence of exposure to conflicting arguments
\citep{Taber2006motivated,Taber2009motivated,Lord1979biased,McHoskey1995case}, other studies report no evidence 
\citep[e.g.][]{Teel2006evidence,Druckman2011framing,Corner2012uncertainty,Shamon2019changing}. Unfortunately, it is difficult to say as to why those empirical studies find mixed evidence on the issue, because the conceptual and methodological heterogeneity applied in the studies does not allow to draw systematical conclusions \citep[cf.][p. 108]{Shamon2019changing}. 
Hence, despite the fact that biased processing has been shown to be a relatively robust cognitive mechanism, empirical evidence on its consequences for attitude change has been ambiguous. 
Our approach takes into account that biased processing may come in degrees ($\beta$) and shows that the question of attitude moderation versus polarization crucially depends on how strongly subjects engage in biased processing.
When subjects are exposed to a balanced mix of pro and con arguments there is a sharp qualitative transition from attitude moderation to attitude polarization when $\beta$ crosses a critical value.
That is, attitude polarization at the individual level requires a sufficient level of biased processing. 

Secondly, 
the close connection of experiment and theoretical model advanced in this paper provides a method to assess the strength of biased processing $\beta$ from experimental data. 
This is important since biased processing might come in different degrees across different issues. 
Indeed, Shamon et al.'s experiment \citep{Shamon2019changing} addresses attitudes towards
six different technologies for \revision{electricity} 
production and we find a clear signature of biased processing in all of them.
However, there are differences in strength: while attitude data on gas and biomass shows a weak bias clearly below the critical point between attitude moderation and polarization, arguments on coal, wind (onshore and offshore), and solar power are subject to stronger biases above the critical value.
The refined version of ACT would suggest that a group deliberation on one of the former two technologies is less prone to yield dissent compared to the latter four topics.
The approach hence allows to calibrate the microscopic mechanisms of argument exchange employed in ACT with respect to the specific topic addressed in an balanced--argument persuasion experiment. It enables a systematic approach of experiment and theoretical refinement.

\revision{
Finally,}  
incorporated into a computational theory of group deliberation such as ACT, we can address the implications of biased processing at the collective level of groups or populations. Previous modeling work incorporating biased processing has shown that biased assimilation coupled with homophily may generate patterns of collective polarization if the bias is sufficiently strong \citep{Dandekar2013biased}. \cite{Dandekar2013biased} model biased processing in such a way that it "mathematically reproduces the empirical findings of \cite{Lord1979biased}" \citep[][p. 5793]{Dandekar2013biased}. 
However, they miss to describe the theoretical micro process that underlies information processing as well as resulting attitude changes in detail and conclude that homophily alone is not sufficient for polarization. This is in disagreement with one of the main results of ACT \citep{Maes2013differentiation} which demonstrated that homophily alone may explain polarization under positive social influence with unbiased argument adoption. We integrate biased processing into the framework of ACT to obtain a clearer picture of its collective level implications. We show that weak biased processing leads to a very efficient process in which a group jointly supports one alternative whereas strong biased processing leads to an intermediate phase in which two subgroups with strongly opposing views emerge. This bi-polarization phase becomes exponentially more persistent with an increase in processing bias. 
Thus, we show that in the absence of other mechanisms, attitude polarization at the individual level is a prerequisite for collective bi-polarization. Homophily is not necessary but accelerates the polarization process and stabilizes a conflictual, bi-polarized group situation. 


\revision{

The remainder of this paper is structured as follows:
We briefly comment on terminological choices in the next section.
The balanced argument experiment will be described in Section \ref{sec:experiment}.
Section \ref{sec:theoryandmodel} describes how biased processing is integrated into the setting of ACT.
In Section \ref{sec:microlevelimplications}, we will take the perspective of an individual subject and derive the response function for the expected attitude changes.
This is applied to the data from the experiment in Section \ref{sec:calibration}.
The macro-level implications of biased processing will be studied in Section \ref{sec:macroimplications}.
Finally, Section \ref{sec:discussion} summarises the main contributions of the paper, and discusses its limitations as well as future directions.

\section{Terminology}

Given that our research design combines psychological research on attitudes with opinion dynamics models a brief note on terminology  
might be helpful: Throughout this paper, we will use the terms attitude and opinion interchangeably. We would like to embrace that attitude is more established in the context of persuasion experiments and opinion is the more typical term in opinion dynamics research.
While the main focus of the former is on individual attitude change, the latter is mainly concerned with collective processes modeling groups or large populations. In the context of polarization, this may lead to confusion because the psychological concept of attitude polarization and the sociological theorizing behind opinion polarization relate to rather different phenomena. In the first case, attitude polarization relates to the persuasive effect that the attitude of a single individual becomes more extreme in either direction after an informational treatment.
In the second case, opinion polarization refers to a bi-modal distribution of opinions in a population and to the dynamical process by which such a distribution emerges \citep[cf.][p. 693]{DiMaggio1996have}.

For the purposes of the paper we distinguish two qualitatively different patterns of attitude change at the individual level (see Figure \ref{fig:terminology}). 
Consider that an attitude is measured before (blue point) and after (red point) a treatment. If the attitude is less strong and approaches the neutral point after the treatment, this is called attitude moderation. 
Conversely, if the initial attitude is reinforced and more extreme after the treatment, this is called attitude polarization.
At the collective level, we consider how an initial distribution of opinions in a population (blue) evolves in repeated interactions.
As exemplified by the red curves in Figure~\ref{fig:terminology}), we differentiate three qualitatively different outcomes that will be relevant in the analysis that follows: moderate consensus (bottom left), extreme consensus (middle), or opinion polarization (left).


\begin{figure}[ht]
	\centering
	\includegraphics[width=0.95\linewidth]{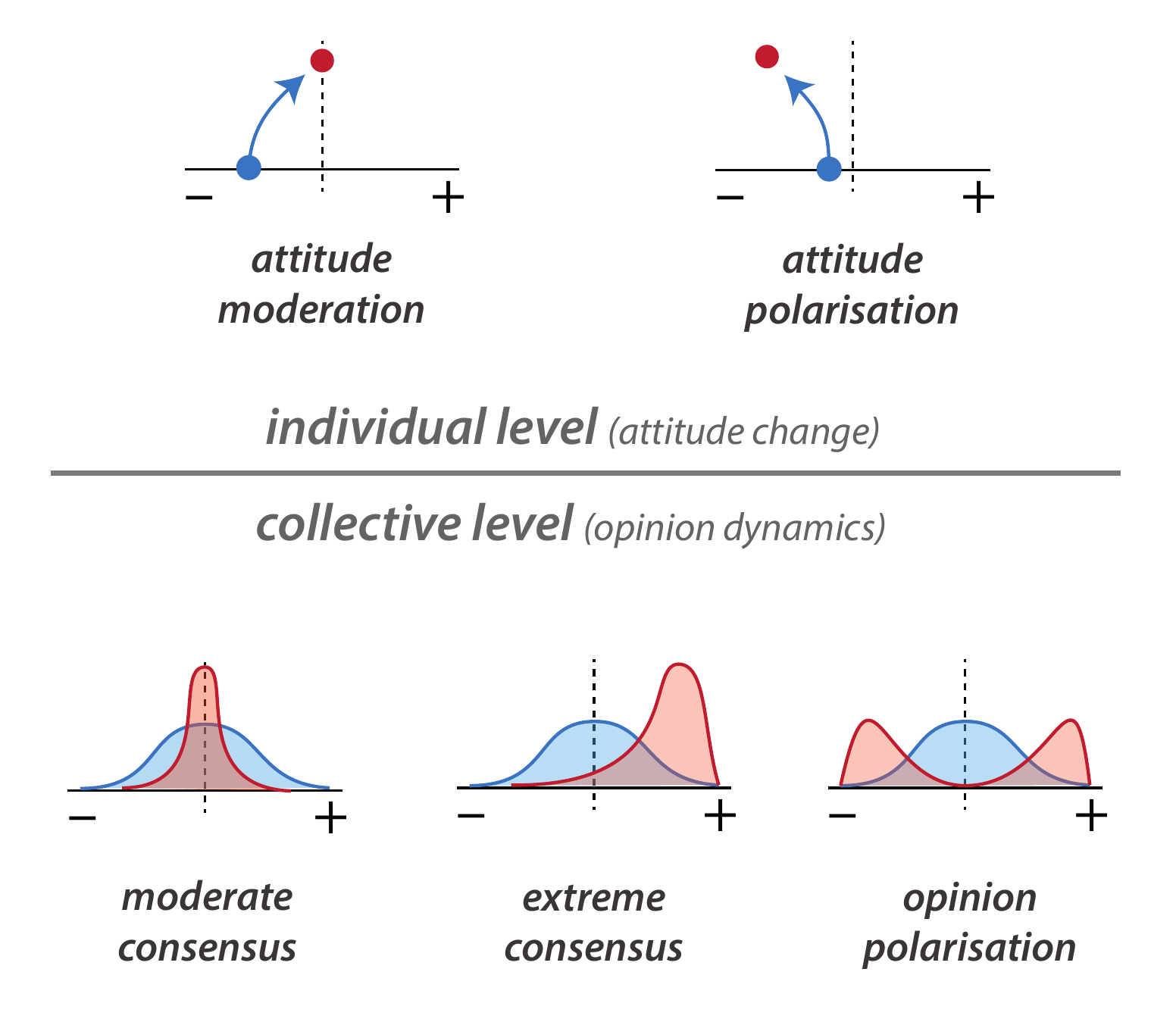}
	\caption{Overview of individual and collective phenomena discussed in the paper.
	}
	\label{fig:terminology}
\end{figure}

}

\section{Experiment}
\label{sec:experiment}

\revisionHS{In 2017, \citep{Shamon2019changing} designed an online survey experiment to assess the impact of biased processing of arguments on attitude change regarding different electricity generating technologies. Participants were recruited from a voluntary-opt-in panel of a non-commercial German access panel operator.\footnote{Participants were neither paid nor promised any monetary incentives, but presented with the prospect of receiving a summary of the survey results upon request.} The external validity of the study's empirical findings is limited since voluntary-opt-in panels suffer from self-selection bias due to the (non-probabilistic) recruitment process of panel lists. Furthermore, the non-commercial access panel operator did not provide the option of using a quota sampling procedure for the survey experiment. This implies that none of the respondent characteristics in the sample necessarily matches the distribution in the population by design.\footnote{ Nevertheless, (non-commercial) access panels are increasingly being used for conducting online survey experiments such as for factorial surveys in social science \citep[e.g.][]{Sattler2021why,Shamon2022factorial}.
} 


\revisionHS{The analytical sample consists of 1,078 persons who indicated to have a residential address (principal address) in Germany. Respondents' average age in the analytical sample is 40.8 years (SD=15.7), and 49.3 percent of respondents are female, 49.4 percent are male, and 1.3 percent refused to classify their gender. Furthermore, 77.7 percent of the respondents had received a secondary school leaving certificate and 5.3 percent stated that they are employed in the energy sector.} 

In the experiment, respondents' attitudes towards six technologies were measured\footnote{Initial and posterior attitudes towards electricity generating technologies were measured on a nine-point scale (0: strongly against the technology; 4: neither against nor in favor of the technology; 8: strongly in favor of the technology), whereas respondents were also offered an exit option (cannot choose).} both before (initial attitudes) and after (posterior attitudes) the presentation of 14 arguments on one of six technologies (Setting 1: coal power stations; Setting 2: gas power stations; Setting 3: wind power stations (onshore); Setting 4: wind power stations (offshore); Setting 5: open-space photovoltaic; Setting 6: biomass power plants).\footnote{The 84 (=14*6) arguments were developed by an interdisciplinary expert team consisting of engineers and physicists, economists, and social scientists at the Institute of Energy and Climate Research – Systems Analysis and Technology Evaluation (IEK-STE) at Forschungszentrum Jülich.} 

\revisionHS{Respondents were randomly assigned to only one of the six settings.} The set of arguments was balanced in the sense that it comprised seven arguments speaking in favor (pro arguments) and seven arguments speaking in disfavor (counter arguments) of the respective technology. \revisionHS{Each argument was presented on a separate page of the online questionnaire. In order to prevent response-order effects, the order of the argument blocks (block of pro arguments followed by a block of counter arguments vs. a block of counter arguments followed by a block of pro-arguments) as well as the order of arguments within each block was randomized.}

Respondents were asked to rate each argument's persuasiveness as well as to state their perceived familiarity with each argument.  
The research design allowed to assess not only to what extent initial attitudes affect persuasiveness ratings of arguments but also to what extent respondents' initial attitudes change after the exposure to the balanced set of 14 arguments.

\revisionHS{Respondents' persuasiveness ratings were registered for each argument on a nine-point scale (0: the argument is not at all persuasive; 8: the argument is very persuasive).
Next to the persuasiveness rating scale, respondents could state their perceived familiarity with each of the 14 arguments (0: I am not aware of this argument; 1: I am aware of this argument). This allowed to calculate a balance of argument ratings for each respondent. The balance of argument ratings was calculated by subtracting a respondent's average persuasiveness rating for the seven counter arguments from the
average persuasiveness rating for the seven pro arguments. Hence, a persuasiveness balance ranges from -8 (meaning that a respondent rated all seven counter arguments with 8 while rating all pro arguments with 0) to +8 (meaning that a respondent rated all seven pro arguments with 8 while rating all counter arguments with 0). The balance of ratings serves as a proxy variable for respondents' engagement in biased processing}.}

\begin{figure}
    \centering
    \includegraphics[width=0.85\linewidth]{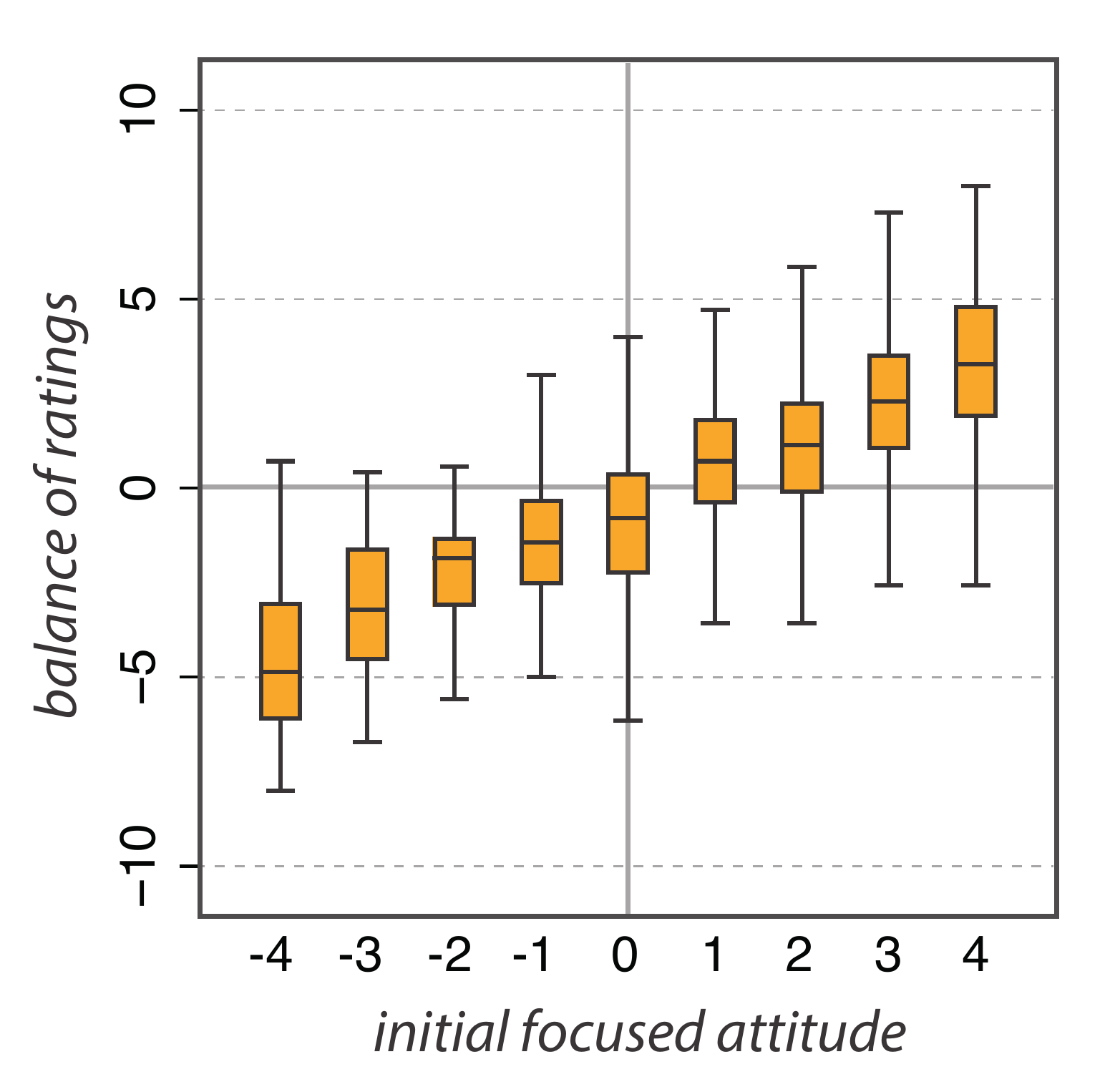}
    \caption{Balance of argument ratings as a function of the initial focused attitude.}
    \label{fig:persuasiveness}
\end{figure}

The experiment provides empirical evidence that persons' engagement in biased processing depends systematically on their initial attitude. Figure \ref{fig:persuasiveness} shows \revisionHS{for each attitude position and across all technologies a box plot (without outliers) of  respondents' balance of argument ratings.} 
\revisionHS{The distribution of respondents' balance of argument ratings depends on respondents' initial attitude towards the respective technology that was focused in the 14 arguments (hereinafter referred to as focused attitudes).} The majority of respondents with initial negative focused attitudes rated counter arguments as more persuasive than pro arguments and the majority of respondents with initial positive focused attitudes rated pro arguments as more persuasive than counter arguments. 
This pattern is perfectly in line with theoretical considerations on biased processing according to which persons tend to inflate the quality of those arguments that conform to their initial attitude and deflate the quality of those arguments that do not conform their initial attitude. 
Among persons with an initial negative as well as persons with an initial positive focused attitude, the persuasiveness balance is biggest in absolute terms at the extreme points of the attitude scale while it is modest among respondents with an initial neutral attitude. Hence, \cite{Shamon2019changing} conclude that respondents process arguments biasedly and their engagement in biased processing increases with the extremity of their attitudes. 

While the subjective ratings of argument persuasiveness confirm systematic biases in the evaluation of arguments, it is of great practical concern how the actual attitudes change after exposure to a balanced set of arguments not clearly in favor or against a certain issue. If attitudes become generally more extreme after exposure to balanced information, the use of arguments  in a societal debate would likely broaden the gap between supporters and opponents of different energy technologies \citep[cf.][]{Shamon2019changing}. For this reason, a lot of experimental research has been invested on answering the question whether biased processing implies attitude polarization when subjects are exposed to conflicting arguments but cannot easily be answered on the basis of empirical evidence due to the conceptual and methodological heterogeneity (see Introduction). 

In order to obtain a more nuanced picture of attitude change under conflicting arguments \cite{Shamon2019changing} suggest to consider dynamics at the individual level by examining transition probabilities conditioned on the initial focused attitude. 
That is, the patterns of attitude change are considered independently for subjects with a negative, a neutral and a positive initial attitude. Induced attitude changes, in turn, are categorized with respect to polarization (more extreme), persistence (unchanged) and moderation (less extreme). 
This reveals that both attitude polarization and moderation may occur simultaneously at the individual level and that these effects may average out at the aggregate level of the entire population. 
While the analysis in \cite{Shamon2019changing} allows for a more fine-grained understanding of the role of attitude extremity and its impact on biased processing,  
it still remains puzzling what degree of biased processing is required for the emergence of attitude polarization. 

In this paper we bring the analysis of attitude-dependent attitude changes to a higher level of sophistication by deriving a statistical model for the full distribution of conditional attitude change based on cognitive principles.
This allows us to vary the strength of biased processing and to determine how well empirically observed attitude changes are matched by a specific value.
Starting from the cognitive structure that underlies ACT, we incorporate biased argument adoption and analyze the attitude changes that would be expected under the given experimental conditions (exposure to a balanced set of arguments).
We account for the strength of biased processing by a parameter $\beta$ which governs the extent to which evaluation biases (Figure \ref{fig:persuasiveness}) lead to biases in argument adoption.
This makes explicit, among other things, that attitude persistence at the extreme ends of the attitude scale is indicative of a rather strong processing bias contributing to a global pattern of attitude polarization. 
We show that there is a sharp transition from attitude moderation to polarization as $\beta$ increases rendering in-principle statements that biased processing leads to attitude polarization somewhat ill-posed. 
Most importantly, as the processing bias $\beta$ may depend on the issue under investigation one can expect attitude moderation in some and polarization in other cases.

\section{A cognitive model of biased argument processing}
\label{sec:theoryandmodel}

\subsection{Attitude structure}

\begin{figure}[ht]
	\centering
	\includegraphics[width=0.95\linewidth]{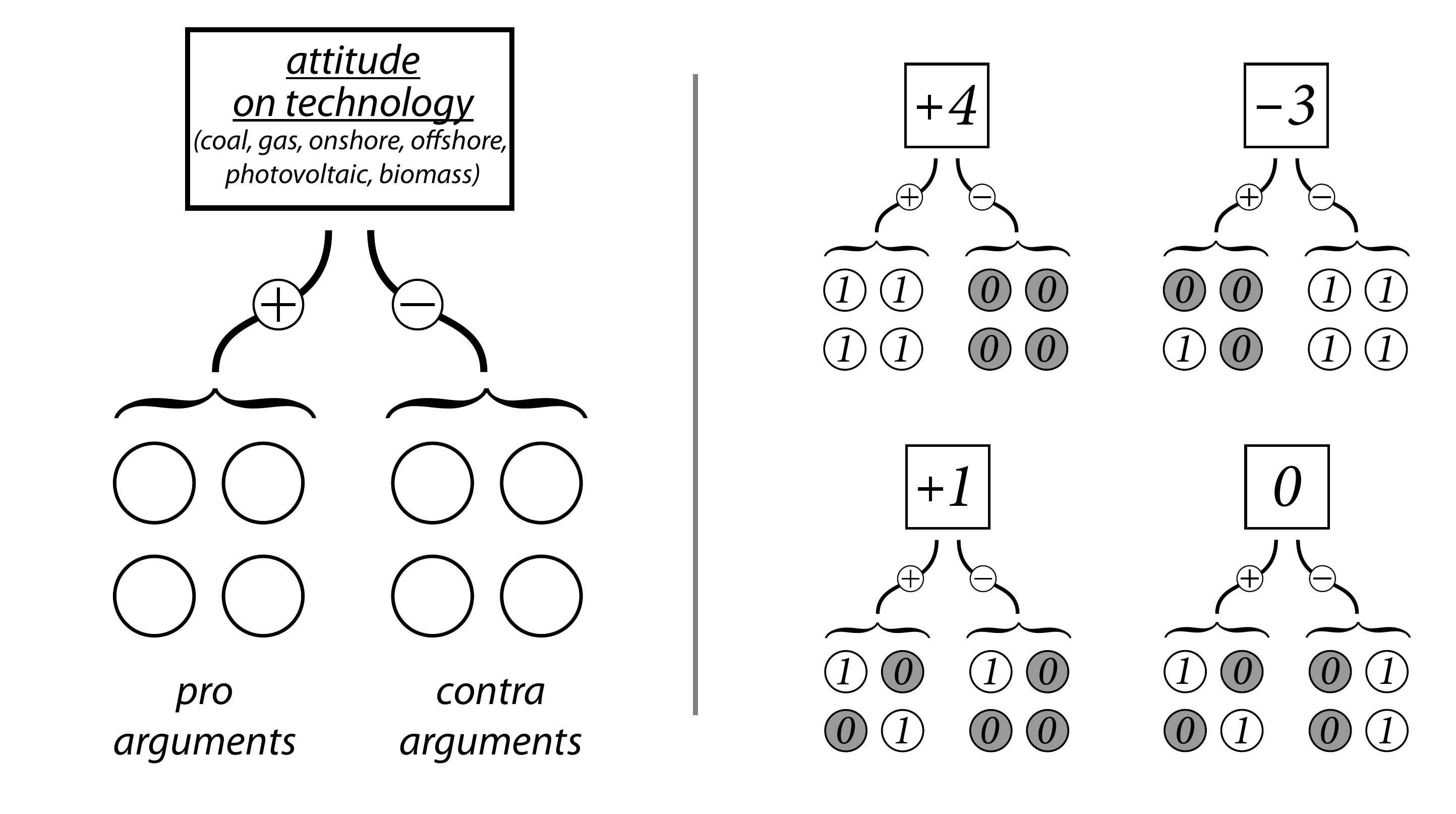}
	\caption{Structure of opinions assumed by ACT (left) and four example configurations (right).
		Sets of pro- and con-arguments are assumed to underlie the attitudes towards different issues (energy production technologies).
		Single arguments can either be believed (1) or not (0). 
		The numbers of pro- and con-arguments that an agent beliefs in determine the attitude towards the focus issue.
	}
	\label{fig:opinionstructure}
\end{figure}

While the majority of computational opinion models treats the opinion as an atomic unit, argument-based models \citep{Maes2013differentiation,Banisch2021argument} operate with a representation of opinions that takes some degree of cognitive complexity into account.
Individuals usually hold concrete or abstract beliefs on attitude objects that imply a  positive or negative evaluation of the attitude object and form an important basis for attitudes \citep{Eagly1993psychology}. The extent to which positive (or negative) connoted beliefs outweigh negative (or positive) beliefs on an attitude object in an individual's belief system, determines in tendency the valence (positive or negative) and extremity of a person's attitude. Hence, ignoring this formative structure of attitudes may lead to the fact that essential mechanisms cannot be identified.
In ACT, this is modeled by a set of arguments that support either a positive (pro) or a negative standing (con) towards the issue at question.
Agents can either believe and therefore adopt an argument or reject it and the net number of pro- and con-arguments determines the overall attitudinal judgment.
That is, an agent's attitude towards an issue (an electricity production technology in our case) is positive to the extent to which the number of pro-arguments exceeds the number of con-arguments in its belief system.
This setting is shown in Figure \ref{fig:opinionstructure} along with four example argument configurations and the respective attitude.

Formally, let us denote the number of possible pro- and con-arguments by $N_{+}$ and $N_{-}$ respectively.
We denote a single argument by $a_i$ where $i$ is used to index the set of all arguments.
Following earlier models \citep{Maes2013differentiation,Banisch2021argument}, we assume that only two values are possible for each argument: $a_i = 1$ indicates that the argument is believed, and $a_i = 0$ that it is rejected. 
We further denote by $e_i$ the evaluative contribution of argument $i$ to the attitude. 
The $e_i$'s are one for pro-arguments and minus one for con-arguments.
An agent's opinion $o$ is then given by
\begin{equation}
	o = \sum_{i}^{} a_i e_i = n_{+} - n_{-}
	\label{eq:opinion}
\end{equation}
with $n_{+}, n_{-}$ the number of currently held pro- and con-arguments respectively.
On the right hand side of Figure \ref{fig:opinionstructure} four different argument configurations are shown along with the resulting opinion for a setting with $N_{+} = N_{-} = 4$.
Maximal support ($o=+4$) is obtained when agents believe in all pro- and no con-argument.
A maximally negative opinion ($o=-4$) means that all con-arguments are considered valid and pro-arguments are rejected.
Equation \ref{eq:opinion} hence leads to opinions on a nine-point scale from -4 to +4, in agreement with the attitude scale used in the experiment.

\subsection{Biased argument evaluation}

The experiment described in \cite{Shamon2019changing} has revealed a linear relationship between the current attitude and the evaluation of argument persuasiveness (see Figure \ref{fig:persuasiveness}).
One explanation for this phenomenon is that persons with a positive or negative attitude show the motivation to produce defensive responses to attitude incompatible arguments while they are motivated to develop favorable thoughts on attitude-consistent arguments \citep{Kunda1990case,Petty1986elaboration}.
Another explanation for this argument evaluation bias might be seen in individuals striving for cognitive coherence \citep{Festinger1957theory,Thagard1998coherence}.

To see this, let us regard the attitude structure described above as a simple cognitive network comprised of beliefs and a single attitude node which are linked by evaluative associations.
We can define the coherence of a cognitive configuration made up by a specific argument string $a$ and an opinion $o$ by the net number of attitude-coherent versus attitude-incoherent evaluative associations weighted by attitude strength
\begin{equation}
C(a,o) = \frac{1}{2} \sum_{i} (2a_i-1) e_i o = \frac{1}{2} \sum_{i} (2a_i-1) e_i (n_{+} - n_{-}).
\end{equation}
The transformation $(2a_i -1)$ leading from $\{0,1\}$ to $\{-1,1\}$ is introduced because we want to take into account the contribution of rejected arguments $a_i = 0$ and the prefactor $1/2$ is introduced for the respective normalization.
If an agent is exposed to a new argument $a_i'$, we assume that the evaluation $V(a_i')$ of it depends on whether $a_i'$ leads to an increase or decrease in cognitive coherence, that is, on the difference $C(a',o)-C(a,o)$.
For the opinion structure described above this yields 
\begin{equation}
V(a_i')  = C(a',o)-C(a,o) = (a_i' - a_i) e_i (n_{+} - n_{-}).
\label{eq:argumentevaluation}
\end{equation}
In other words, the evaluation of a new pro-argument (i.e. $e_i = 1$) is a linear function of the current opinion with $V(a_i') = o = (n_{+} - n_{-})$.
A new counter argument ($e_i = -1$), conversely, is evaluated as $V(a_i') = -o = (n_{-} - n_{+})$.
This aligns well with the linear relationship between initial attitude and bias in the rating of arguments that has been identified in the experiment (Figure \ref{fig:persuasiveness}).

\subsection{Biased argument adoption}

If an agent is exposed to a new argument ($a_i' \in \{0,1\}$) either from peers (Section \ref{sec:macroimplications}) or in an experimental treatment (Section \ref{sec:microlevelimplications}) (s)he may adopt the argument or not.
In current implementations of ACT \citep{Maes2013differentiation,Feliciani2020persuasion,Banisch2021argument} that do not incorporate intra-personal processing biases, this adoption probability (denoted as $p$) is homogeneous and independent of the current attitude ($p = 1/2$).
Biased processing posits that the probability of argument assimilation depends on the current opinion (i.e. $p(o)$) in such a way that attitude-coherent arguments are adopted with a high probability ($p(o) > 1/2$) whereas this probability is reduced if the argument is incoherent with the opinion ($p(o)< 1/2$).

We are, however, not rational optimizers of cognitive coherence but largely unconscious processes drive changes in our cognitive system.
It would be highly implausible to assume that individuals with a negative attitude will never accept a pro argument.
Biased processing as conceptualized here in terms of a strive for cognitive coherence comes in degrees.
To take this into account we introduce a free parameter $\beta$ for the \emph{strength of biased processing} which determines the extent to which congruent arguments are favored over incongruent ones.
We use the logistic sigmoid function
\begin{equation}
p_{\beta}(V(a_i')) = \frac{1}{1 + e^{- \beta V(a_i')}}
\label{eq:argadoptionmodel}
\end{equation}
as a probabilistic model in which the probability to adopt or reject a new argument depends on the evaluation $V(a_i')$ of that argument in a non-linear way.
For further convenience, we shall differentiate the cases that $a_i'$ is a pro- or a con-argument and rewrite 
\begin{eqnarray}
p_{\beta}^+(n_+,n_-) = \frac{1}{1 + e^{\beta (n_{-} - n_{+})}}\nonumber\\
p_{\beta}^-(n_+,n_-) = \frac{1}{1 + e^{\beta (n_{+} - n_{-})}}
\label{eq:argadoptionmodel2}
\end{eqnarray}
which takes into account the linear relationship between argument evaluation $V(a_i')$ and attitude $o = (n_{+}- n_{-})$.

\begin{figure}[ht]
	\centering
	\includegraphics[width=0.89\linewidth]{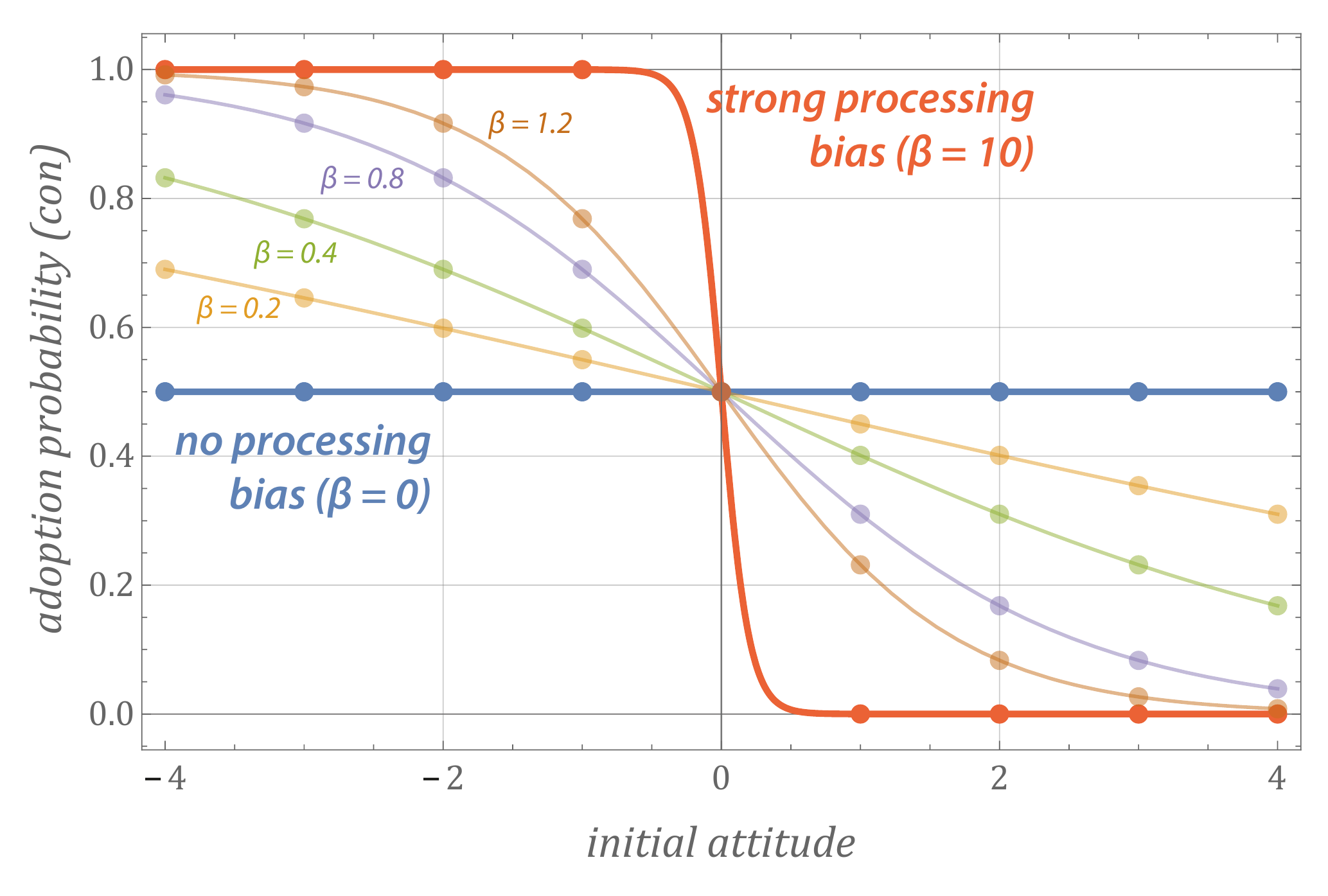}
	\caption{Probability to adopt a con-argument ($p_{\beta}^-$) by as a function of the current attitude for different values of biased processing strength $\beta$.}
	\label{fig:adoptionprobability}
	
\end{figure}

Figure \ref{fig:adoptionprobability} shows the behavior of this probabilistic choice model for the case that an agent is confronted with a con-argument ($p_{\beta}^-(n_p,n_c)$).
Unless $\beta = 0$, the probability of adoption is higher than chance if the current opinion is negative and smaller than 1/2 if it is positive.
If $\beta$ is large (bold orange line), the adoption of incoherent arguments becomes virtually zero and we approach the regime of the rational optimizer. 
If, on the other hand, $\beta$ is zero (bold blue curve), there is no adoption bias and arguments are adopted with a homogeneous probability of 1/2. 
This limiting case therefore corresponds to the choice made in previous ACT models.

\section{Theoretical implications for the balanced-argument treatment}
\label{sec:microlevelimplications}

In this section we take the perspective of an individual subject.
Using the cognitive model of attitude-dependent biased processing described in the previous section, we derive a subject's expected reactions after exposure to an unbiased set of arguments.
This allows a precise characterization of whether attitude polarization or moderation is expected at the individual level by exposure to conflicting arguments as realized in the experiment (see \cite{Shamon2019changing} and Section \ref{sec:experiment}).

\subsection{Expected attitude change after exposure to an unbiased set of arguments}

In the experiment (see Section \ref{sec:experiment}), subjects are confronted with an unbiased set of pro- and con-arguments.
Attitudes are measured before and after the treatment and the effect on attitude change is analyzed.
In order to relate these experimental findings to the microscopic assumptions about argument exchange in the model, we ask: How would artificial cognitive agents react to the same experimental treatment and what is their expected attitude change?
For this purpose, we consider that the opinion structure is comprised of four pro- and con-arguments respectively (see Figure \ref{fig:opinionstructure}).
For further convenience, we shall denote this number by $M = N_{+} = N_{-} = 4$.
Each pro-argument, if believed, contributes with +1 to a positive attitude, each con-argument with -1 to a negative stance and we have chosen this setup to align the model with the experiment in the sense that attitudes lie on a 9-point scale ranging from -4 to +4.

Let us assume that an agent receives an unbiased set of four pro- and four con-arguments at once. 
Attitude change may only take place if at least one $a_i'$ is new to the agent and if it is adopted.
That is, it depends in two different ways on the number of currently held pro- and con-arguments ($n_{+}$ and $n_{-}$ respectively).
First, with a certain probability arguments are already shared by the agent ($a_i = a_i'$) and do not present new information.
Second, as $n_{+}$ and $n_{-}$ define the current attitude of the agent by $o = n_{+} - n_{-}$ they are relevant for the biased adoption probabilities $p_{\beta}^+$ and $p_{\beta}^-$ defined in (\ref{eq:argadoptionmodel2}).
Namely, for an agent that already believes in $n_{+}$ pro-arguments, the probability of adopting $k$ additional pro-arguments is given by the binomial distribution
\begin{equation}
Pr_{n_{+}}[\Delta n_{+} = k] = (p_{\beta}^+)^k  (1-p_{\beta}^+)^{N_{+} - n_{+} -k} \binom{N_{+}-n_{+}}{k}
\label{eq:AdoptPro}
\end{equation}
where the adoption probability $p_{\beta}^+$ depends on $n_{+}$ and $n_{-}$ as given by Eq. (\ref{eq:argadoptionmodel2}).
For con-arguments, we have equivalently
\begin{equation}
Pr_{n_{-}}[\Delta n_{-} = k] = (p_{\beta}^-)^k  (1-p_{\beta}^-)^{N_{-} - n_{-} -k} \binom{N_{-}-n_{-}}{k}.
\label{eq:AdoptCon}
\end{equation}
Notice that an attitude change of $k$ implies that the difference between adopted pro- and con-arguments is exactly $k$.
Consequently, the probability that an attitude change of $k$ is observed after exposure to all arguments is given by
\begin{equation}
Pr_{n_{+},n_{-}}[\Delta o = k] = \sum\limits_{l=k}^{M} Pr_{n_{+}}[\Delta n_{+} = l]Pr_{n_{-}}[\Delta n_{-} = l-k].
\label{eq:ProbAttitudeChange}
\end{equation}

Eq. (\ref{eq:ProbAttitudeChange}) completely characterizes the distribution of attitude changes conditioned on the numbers of currently held pro- and con-arguments.
On its basis, the mean attitude change for an agent with $n_{+}$ pro- and $n_{-}$ con-arguments can be computed and is given by
\begin{equation}
\mathbb{E}[\Delta o \arrowvert n_{+},n_{-}] = \sum\limits_{k = -M}^{M} k Pr_{n_{+},n_{-}}[\Delta o = k].
\label{eq:MeanAttitudeChange}
\end{equation}
Notice that the mean expected attitude change $\mathbb{E}[\Delta o \arrowvert n_{+},n_{-}]$ depends on $n_{+}$ and $n_{-}$ and is not equal for all configurations $(n_{+},n_{-})$ that give rise to the same attitude $o$ except for the trivial case of $\beta = 0$.
For instance, an opinion $o = 0$ may result from $(n_{+},n_{-}) = (0,0)$ or $(n_{+},n_{-}) = (4,4)$.
While the probability that the argument treatment presents new previously rejected arguments to the agent is large in the first case, it is zero in the latter.
Since we do not in general know whether an opinion $o$ came about by one or another argument configuration 
we may assume that all argument configurations are equally likely (maximum entropy assumption).
With this assumption, the expected attitude change $\Delta o$ conditioned on the initial attitude $o$ can be written as
\begin{equation}
\mathbb{E}[\Delta o \arrowvert o] = 
2 \tanh \left(\frac{\beta o}{2}\right)-\frac{2 o}{M}
\label{eq:expectedchange}
\end{equation}
where $M$ is the number of pro- and con-arguments.

Eq. (\ref{eq:expectedchange}) characterizes how agents endowed with the cognitive model described in Section \ref{sec:theoryandmodel} would react on average when exposed to an unbiased set of arguments.
The artificial treatment 
for which it has been derived was designed to establish correspondence with the actual treatment in the experiment. 
We will use this relation to assess the strength of biased processing in the context of energy production technologies in Section \ref{sec:calibration}.
However, the model also provides more general insight into whether attitude moderation or polarization is expected after exposure to balanced arguments and may hence provide a new perspective on the mixed empirical evidence on that question \citep{Lord1979biased,Taber2006motivated,Taber2009motivated,Druckman2011framing,Corner2012uncertainty,Teel2006evidence,Shamon2019changing}.


\subsection{Attitude moderation versus polarization}
\label{sec:moderationpolarization}

\begin{figure}[ht]
	\centering
	\includegraphics[width=0.89\linewidth]{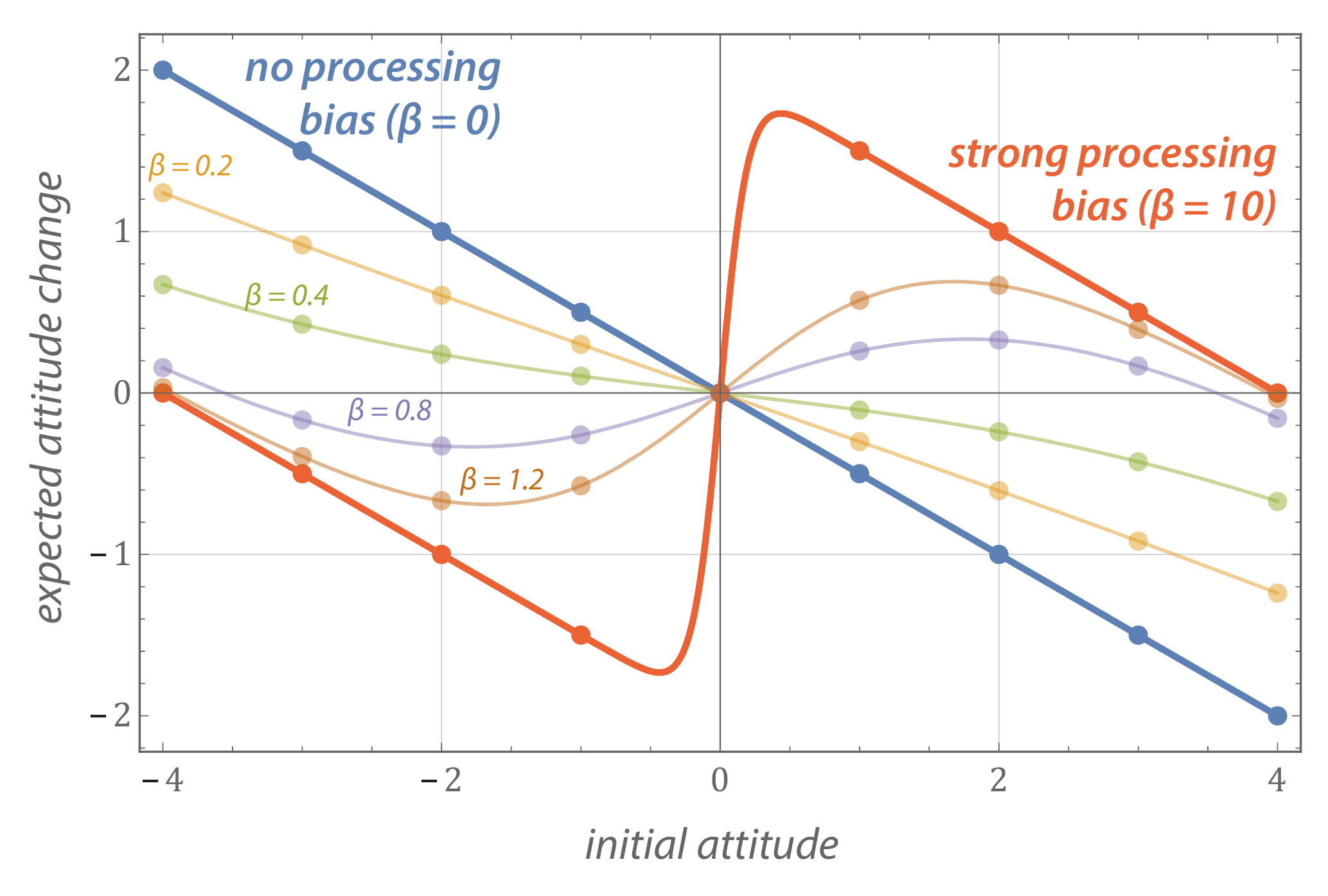}
	\caption{Expected attitude change after exposure to an unbiased set of arguments ($\mathbb{E}[\Delta o \arrowvert o]$) as a function of the current attitude for different values of biased processing strength $\beta$.}
	\label{fig:expectedchange}
\end{figure}

Fig. \ref{fig:expectedchange} shows the behavior of \revision{the response function} (Eq. \ref{eq:expectedchange}) for different values of biased processing $\beta$ as a function of the initial attitude $o$. 
\revision{It provides an overview of the expected attitude change for an individual with a current attitude of $o$.}
As described above, we have used the setting of four pro- and four con-arguments so that $M = 4$ \revision{and the attitude can take discrete values from -4 to 4 as in the experiment.}
With $\beta = 0$ (no bias) (\ref{eq:expectedchange}) reduces to the linear relationship $E_{\beta}(\Delta o \arrowvert o) = -o/2$ which is shown by the blue line in Fig. \ref{fig:expectedchange}.
In this case, the expected change for agents with an initially negative stance ($o < 0$) is positive and agents with positive initial attitudes tend to adopt a less positive opinion after the treatment.
Therefore, models with unbiased argument adoption would predict a relatively strong moderation effect when agents receive an unbiased set of arguments.\footnote{Consider, for instance, an agent with a negative attitude of -4 for which the mean attitude change is +2. 
Such an agent believes in all the 4 con--arguments but in none of the pro-arguments.
When exposed to all 8 arguments, no more con-argument can be adopted but, on average, with $p = 1/2$, one half of the pro-arguments will be adopted leading to an increase of 2 in the attitude.}
\revision{More generally, this regime of attitude moderation is characterized by the fact that the only stable fixed point on the scale at which no further change is expected (i.e., $\mathbb{E}[\Delta o \arrowvert o] = 0$) is the neutral opinion ($o = 0$) in the middle of the scale.}

The other limiting case is marked by $\beta \rightarrow \infty$ which is shown by the orange curves in Fig. \ref{fig:expectedchange} (notice that such a sharp bias is exemplified here by $\beta = 10$).
As shown in Fig. \ref{fig:adoptionprobability}, the adoption probability of attitude--challenging arguments is virtually zero so that an agent with $o = -4$ will adopt none of the pro-arguments in this case.
Likewise, with a moderate inclination towards one side of the attitude scale, a further strengthening of this view is likely so that initial attitudes are reinforced.
\revision{The neutral point at $o = 0$ has become unstable due to the effect of opinion reinforcement leading away from moderate positions, and the most extreme opinions have now become the stable ones where no further change is expected.}
That is, \revision{at the individual level} strong biased processing leads to attitude polarization.

\begin{figure}[ht]
	\centering
	\includegraphics[width=0.89\linewidth]{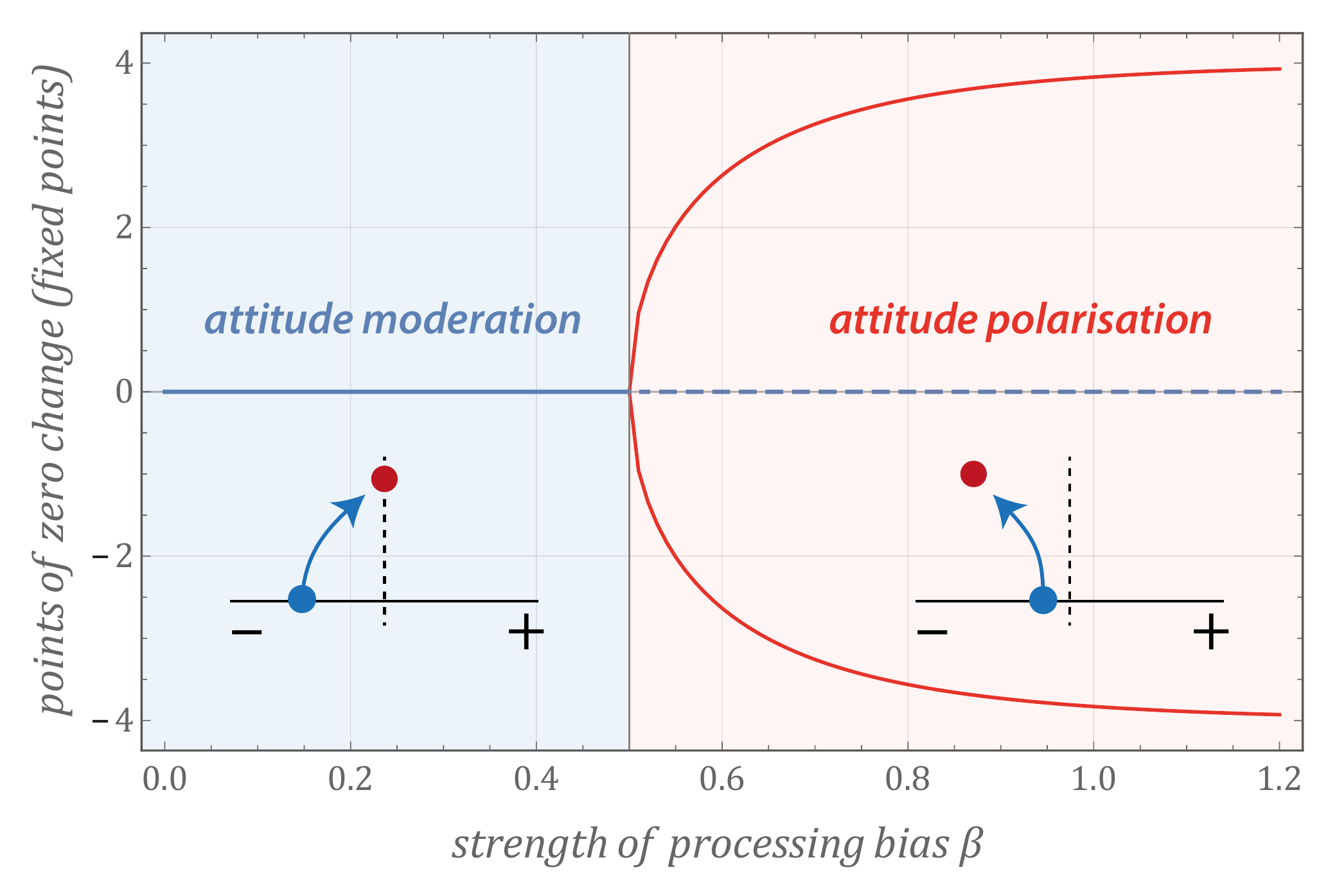}
	\caption{Transition from attitude moderation to attitude polarization as the strength of biased processing ($\beta$) increases. This bifurcation plot shows the attitude values $o$ at which the map $\mathbb{E}[\Delta o \arrowvert o] = 0$ which indicates the no further attitude change is expected for this \revision{value}. The system undergoes a qualitative change from a single stable opinion at $o = 0$ (moderation) to a state where opinions at the extreme end of the attitude scale become stable attracting points (polarization).}
	\label{fig:bifurcationplot}
\end{figure}

This \revision{shows}  
that the puzzle of whether attitude polarization or moderation is likely after exposure to balanced arguments becomes a question of how strong the processing bias is for a given topic of interest.
 \revision{increases from $\beta = 0.4$ (green curve) to $\beta = 0.8$ (violet curve). While a value of $\beta = 0.4$ still leads to moderation, there is a general reinforcement and a polarizing trend with $\beta = 0.8$.
To better understand this qualitative transition in the behavior of the response function we compute a so-called bifurcation diagram, an important tool in dynamical systems theory.
A bifurcation diagram provides a global overview on how the fixed points (i.e., attitudes $o$ which are stable under the treatment, $\mathbb{E}[\Delta o \arrowvert o] = 0$) depend on the relevant parameter ($\beta$).
As shown in Fig. \ref{fig:bifurcationplot}, the response function (\ref{eq:expectedchange}) predicts a relatively sharp transition from attitude moderation to attitude polarization as the strength of biased processing increases beyond a critical value $\beta^* = 1/2$.
}
For $\beta < 1/2$ there is a single fixed point at a neutral attitude of $o = 0$ indicating that individuals tend to moderation. 
As $\beta$ increases the system undergoes a bifurcation in which the neutral fixed point becomes unstable and two stable fixed points at a positive and a negative attitude value emerge.
These two fixed points quickly approach the extreme ends of the attitude scale.
That is, attitudes are attracted towards the extremes after exposure to balanced arguments if biased processing becomes larger than $1/2$.
This corresponds to the regime of attitude polarization.

\section{Experimental calibration}
\label{sec:calibration}

\subsection{Overall assessment}

Eq. (\ref{eq:expectedchange}) can be viewed as a class of statistical models that predict the expected attitude change after balanced argument exposure given an initial opinion.
They are based on the basic assumption of ACT that argument assimilation drives opinion change.
Consequently, the free parameter $\beta$ has a clear meaning in terms of cognitive mechanisms. Namely, it governs to what extent congruent arguments are more likely adopted compared to incongruent arguments.
In this setting we can ask: assuming that agents adapt by argument exchange (as in ACT
models), what is the processing bias $\beta$ that matches best the experimental data on attitude change?
Notice again that previous applications of ACT have not incorporated any bias corresponding to $\beta = 0$. Here we show that biased argument adoption meets better with experimental evidence. 

In order to assess which bias $\beta$ matches best with available experimental data, we compare the theoretical prediction of the cognitive model with the experimentally observed attitude changes by considering the mean squared error (MSE) between (\ref{eq:expectedchange}) and the data.
Let us denote by $(\Delta o_i | o_i)$ the observed attitude change $\Delta o_i$ of subject $i$ after treatment given his or her initial opinion $o_i$.
For an initial attitude $o_i$, the prediction of our model is given by $E_{\beta}(\Delta o | o_i)$ as specified in (\ref{eq:expectedchange}).
The MSE over all observed values is then given by
\begin{equation}
\epsilon_{\beta} = \frac{1}{N_S} \sum\limits_{i = 1}^{N_S} [(\Delta o_i |o_i) - E_{\beta}(\Delta o | o_i)]^2.
\label{eq:MSE}
\end{equation}
In addition, we have used the toolbox for non-linear estimation Stata 14 to find the optimal $\beta$ values which is based on the same error computation.

\begin{figure}[ht]
	\centering
	\includegraphics[width=0.89\linewidth]{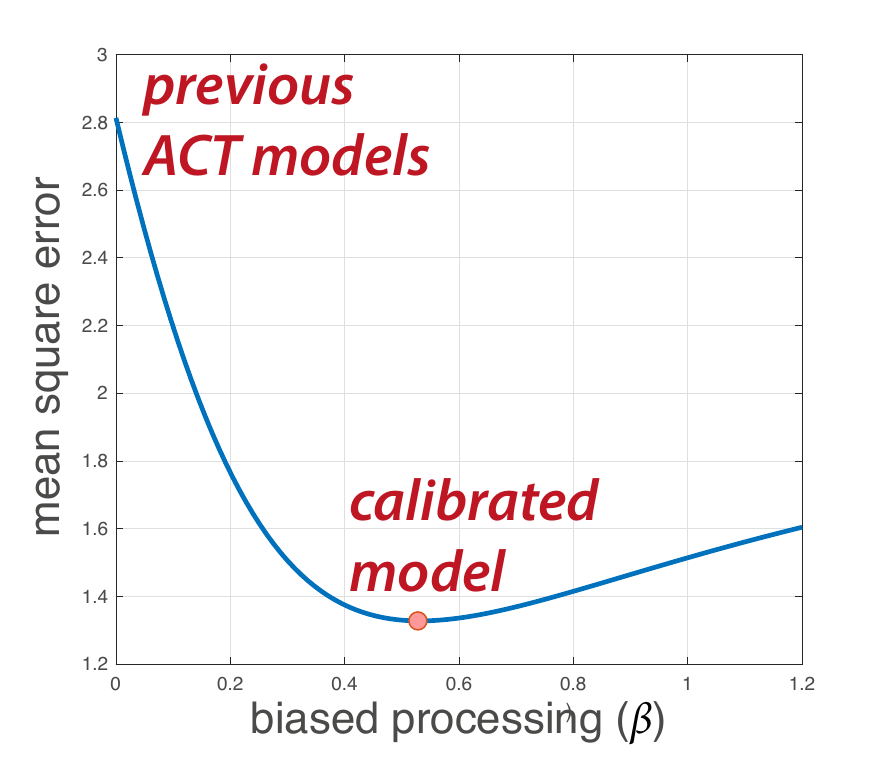}
	\caption{Mean squared error (MSE) between the argument adoption model and the experimental data on attitude change as a function of biased processing strength ($\beta$). The error analysis has been performed on the entire data set ($N = 1078$). As indicated by the red point, moderate biased processing ($\beta \approx 0.5$) meets the data best.}
	\label{fig:overallcalibration}
\end{figure}

In order to identify the optimal $\beta$ we compute the MSE for different values of $\beta$ from zero to 1.2. 
While the former corresponds to unbiased processing, the latter represents a strong processing bias with a clear trend to attitude polarization (see Figure \ref{fig:bifurcationplot}).
For the results shown in Fig. \ref{fig:overallcalibration} 100 equidistant sample points in $[0,1.2]$ are used.

On the whole, we have data on $N_S = 1078$ subjects.
The blue curve in Fig. \ref{fig:overallcalibration} show the MSE $\epsilon_{\beta}$ for the entire data set including all subjects.
The MSE is relatively large for $\beta = 0$, significantly decreases until a minimum value at around $\beta \approx 0.5$ is reached and increases again if $\beta$ becomes larger.
This is a clear indication that the argument adoption process refined with biased processing more appropriately captures argument-induced opinion changes.
A model with moderate biased processing performs significantly better compared to what current implementations of ACT would predict.

\subsection{Differences across issues}
\label{sec:issuedifferences}

\begin{figure}[ht]
	\centering
	\includegraphics[width=0.89\linewidth]{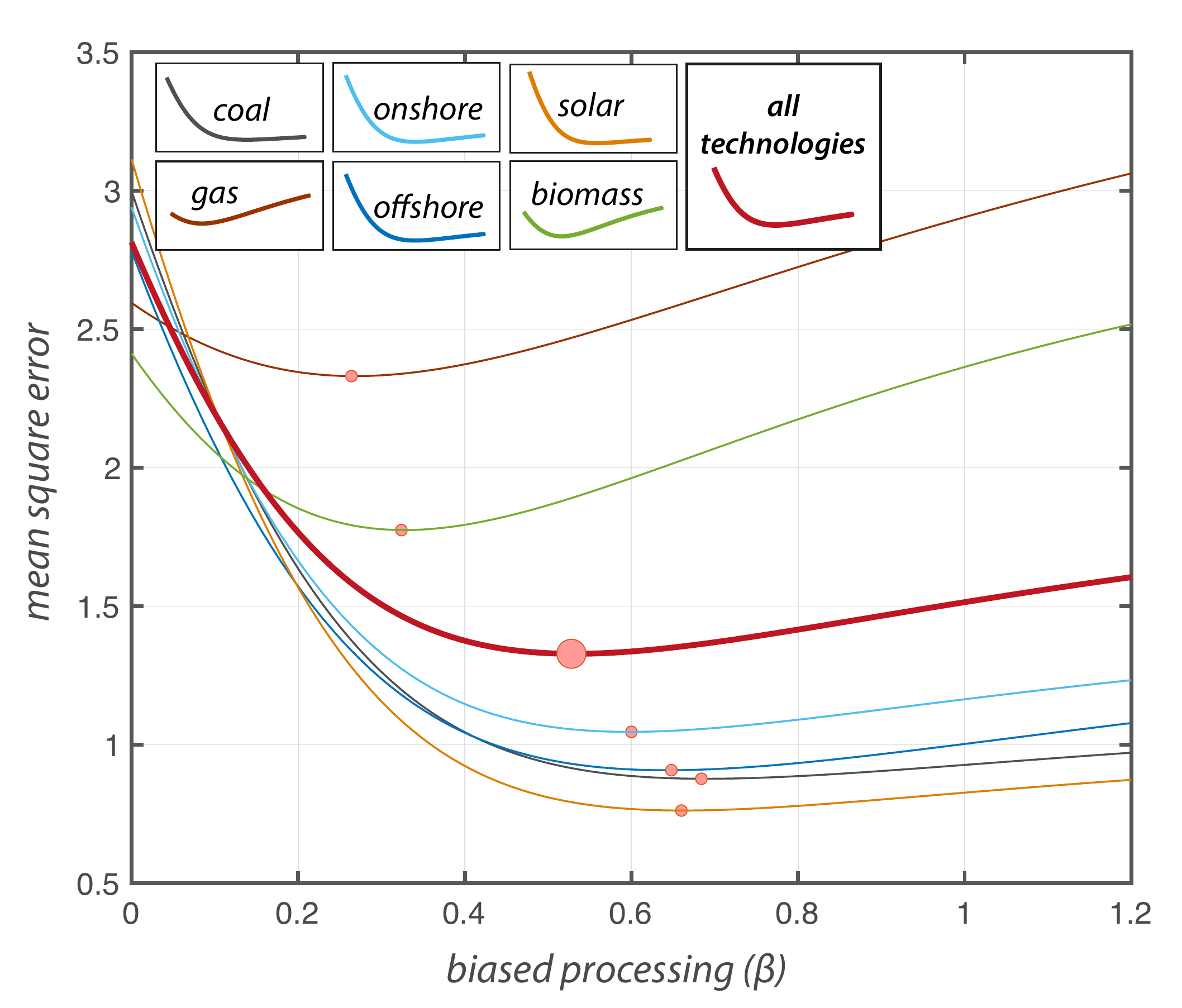}
	\caption{
	Mean squared error between the argument adoption model and the experimental data on attitude change as a function of biased processing strength ($\beta$) for the single technologies. The number of subjects in each technology setting ranges from $N = 170$ (coal) to $N=197$ (solar). While the incorporation of biased processing improves the fit to experimental data in all cases, there are also variations across different energy-generating technologies. Biomass and gas, on the one hand, indicate a level of biased processing clearly below the critical point $\beta^* = 1/2$, the other four provide an optimal fit at values slightly above $\beta^*$.}
	\label{fig:perTechnology}
\end{figure}

Our theoretical considerations in Section \ref{sec:microlevelimplications} have revealed a transition from attitude moderation to polarization when the strength of biased processing crosses a critical value $\beta^* = 1/2$.
This suggests that one reason for the lack of clear evidence for one of the two regimes might be due to variations in the level of biased processing across different topics addressed in the different experiments.
\cite{Shamon2019changing} provides data on six different energy-generating technologies and we can repeat the MSE analysis for each of them independently to gain some insight into these variations.
These results are presented in Fig. \ref{fig:perTechnology}.

This comparison reveals, first of all, a similar qualitative trend for all technologies with an optimal fit at non-zero $\beta$.
However, there are important differences when comparing gas and biomass on the one hand, and coal, wind and solar sources on the other.
First, the processing bias at which attitude change data is matched best is lower for the former. 
Secondly, the error is generally larger for gas and biomass.
As a third observation we notice in Fig. \ref{fig:perTechnology} that for increasing $\beta$ the mean prediction error grows large for gas and biomass, whereas the MSE remains low for strong processing biases in the other cases.
We can only speculate about the reasons for these differences, but they hint at the fact that public discussions on gas and biomass have only recently gained momentum whereas discussions on coal versus wind and solar power have a long history.

Notice that the analysis does not inform us about the "best" model to explain the experimental data.
We have only identified the best model within the class of models defined by $\mathbb{E}_{\beta}$. 
There might be estimators $\mathbb{E}_{\alpha,\beta,\ldots}$ of different form that further reduce the MSE.
On the other hand, the considered model class has been derived from the theoretical assumptions of ACT and has a clear interpretation in this context.
The fact that a specific value $\beta$ of biased processing strength can be found by comparing $\mathbb{E}_{\beta}$ to the data as well as the well-behaved shape of the error curves that render that value as a clearly defined minimum, indicate that a relevant aspect of attitude change processes is captured by this model.
In that sense, the analysis demonstrates that if an argument exchange model is used to analyze collective processes of attitude formation, the microscopic argument adoption process is better aligned with experimental data when a moderate amount of biased processing is incorporated.
We can now turn to the collective-level implications of biased processing.

\section{Collective deliberation with  biased processing}
\label{sec:macroimplications}

Argument communication models describe processes of collective attitude formation as repeated social exchange of arguments.
An artificial population of agents is generated with an initial endowment of random argument strings.
These agents are connected in a social network from which pairs of neighboring agents are drawn at each time step.
One agent acts as a sender $s$ and the second one as a receiver $r$.
The receiver incorporates an argument articulated by $s$ with a probability defined by the cognitive agent model.
While this probability is uniform and independent of the current attitude in previous models it has been refined to incorporate biased processing in this paper.
If a new argument is adopted the attitude of $r$ is updated respectively. 
This process is repeated over and over again until a stable state is reached in which no further change is possible.

Previous work \citep{Maes2013differentiation} has shown that ACT can explain collective bi-polarization if individuals have a strong tendency to interact with similar others (homophily).
In this section, we show that interaction homophily is not necessary for collective bi-polarization.
Biased processing alone can lead to persistent collective states in which one group of agents strongly supports a proposition whereas another group strongly opposes it.

\subsection{Modeling collective argument exchange}

In the model $N$ agents are generated with a random initial assignment of arguments. 
As in the previous sections, we use a setup with 4 pro and 4 con arguments (see Figure \ref{fig:opinionstructure}). 
Consequently, the initial opinion profile is described by a binomial distribution on a 9 point attitude scale. 
In the dynamical process the following steps are performed at each single time step:
\begin{enumerate}
	\item 
	all agents are paired at random ($N/2$ pairs) so that each agent interacts exactly once at each time step (either as sender or receiver),
	\item
	for each pair, the sender $s$ articulates a random argument to the receiver $r$,
	\item
	the receiver adopts that argument with a probability defined by $p_{\beta}$ (Eqs. \ref{eq:argumentevaluation} -- \ref{eq:argadoptionmodel2}), and
	\item
	all agents chosen as receiver in this round update their opinion based on their new argument string. 
\end{enumerate}
After this is done for all pairs of agents, a new round starts with another random pairing of the population.

\begin{figure}[ht]
	\centering
	\includegraphics[width=0.99\linewidth]{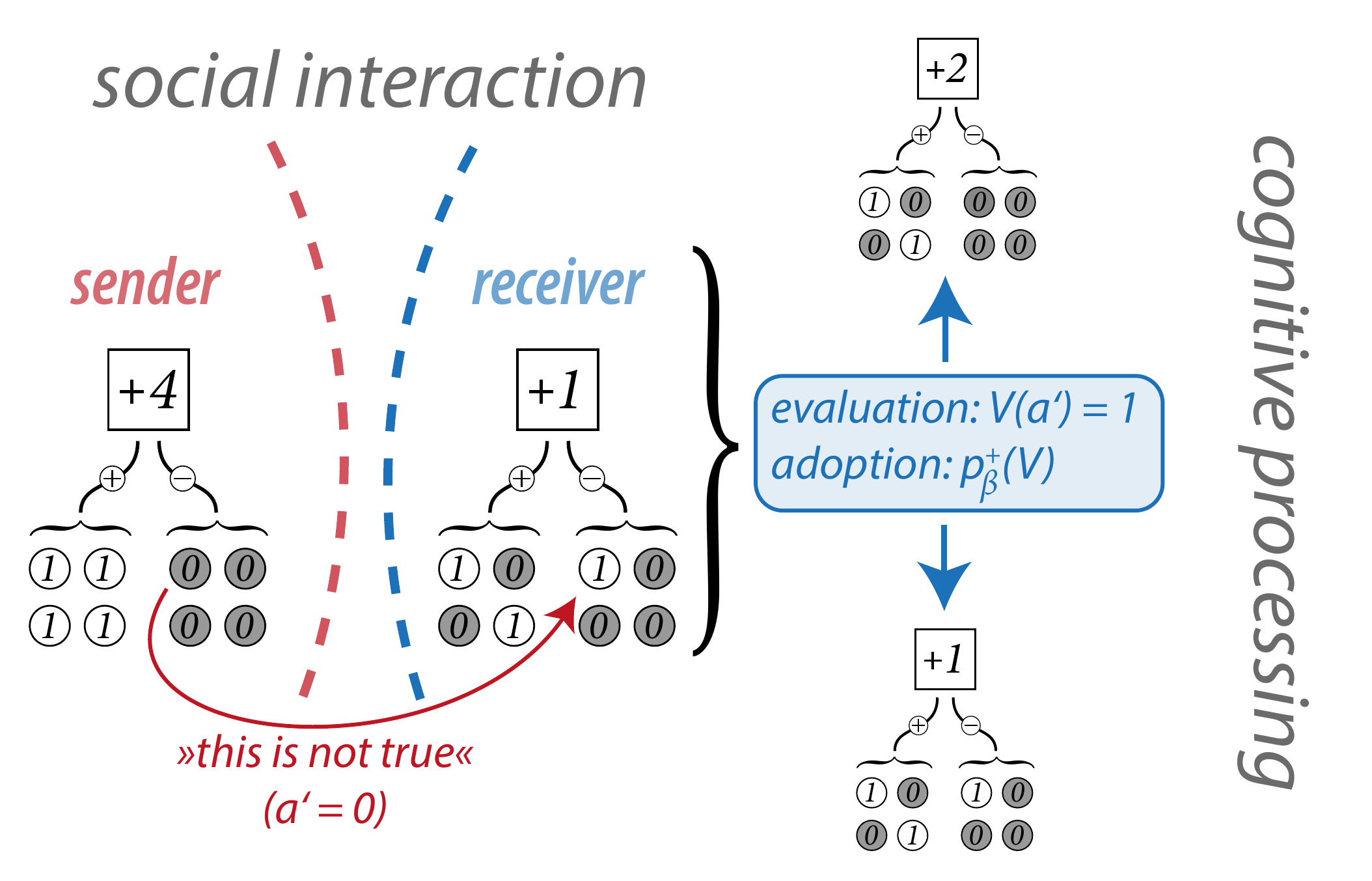}
	\caption{Illustration of the interaction between a sender and a receiver.
	}
	\label{fig:interactionmodel}
\end{figure}

Figure \ref{fig:interactionmodel} illustrates the interaction between the sender and the receiver entering the process with a specific argument set.
The sender is strongly in favor of an issue (e.g. an energy technology) believing in all 4 pro-arguments and rejecting all con-arguments ($n_p = 4$ and $n_c = 0$).
By random selection, $s$ argues for the rejection of one con-argument ($a_i'=0$).
The receiver holding a weak positive attitude currently believes in this argument ($a_i^r = 1$).
By (\ref{eq:argumentevaluation}), $r$ comes to a positive evaluation of $s$'s argument with $V(a_i') = 1$ because it fits with the current attitude $o^r = 1$.
Based on this evaluation, Eq. (\ref{eq:argadoptionmodel2}) decides with which probability $r$ will adopt $a_i' = 0$.
If biased processing is strong, this probability $p_{\beta}$ is close to one and without bias ($\beta = 0$) it is adopted with $p_{\beta} = 1/2$.
Depending on adoption, $r$ either remains with the current opinion or changes towards a slightly more positive view.

\revision{
Notice that our model deviates in an important aspect from the model by \cite{Maes2013differentiation}.
While we assume that agents are aware of all existing arguments but may consider them irrelevant, \cite{Maes2013differentiation} assume that there are many arguments and agents consider a salient subset of them when they form their opinion.
\cite{Banisch2021argument} have shown that the main effect -- bi-polarization in the presence of strong homophily -- is not sensitive to these different choices.
The guiding principle for model development in this paper has been to align as much as possible with the experimental setting which motivated the use of a relatively small set of 4 pro and 4 con arguments.
}

\subsection{Model phenomenology}
\label{sec:modelphenomenology}
	
\begin{figure*}[t]
	\centering
	\includegraphics[width=0.97\linewidth]{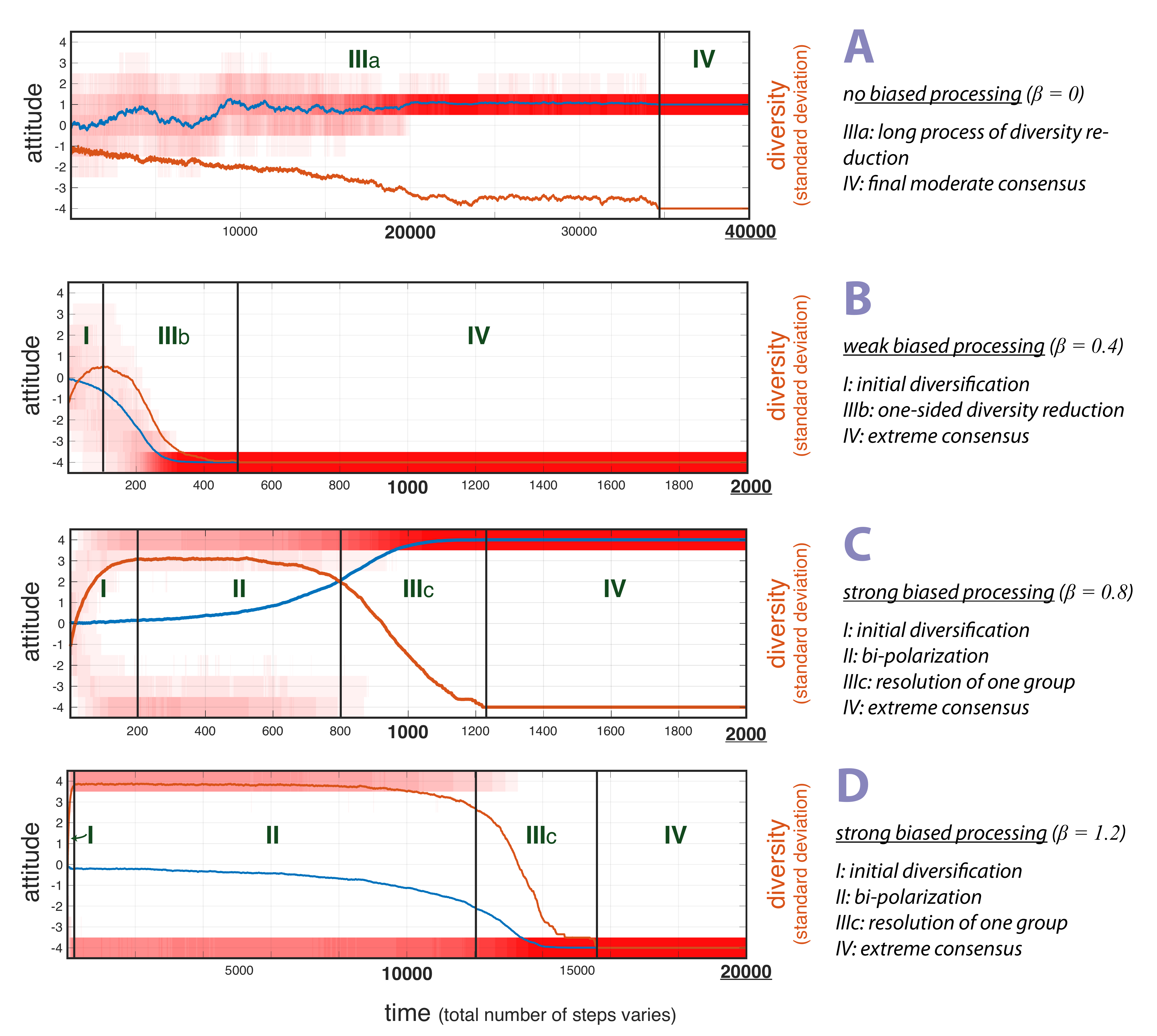}
	\caption{Four paradigmatic model realizations for different levels of biased processing from $\beta = 0$ to $\beta = 1.2$. The figure shows the time evolution of the opinion distribution of a population of $N = 1078$ agents (red shaded on a 9-point attitude scale). It also shows a measure of diversity (standard deviation) and the mean opinion to characterize the distribution. Without biased processing (A) the model leads to a long process in which the population converges to a moderate opinion. Weak biased processing (B, $\beta < \beta^*$) leads to quick global convergence to one or the other side of the attitude scale (choice shift). As biased processing increases (C and D, $\beta > \beta^*$) an intermediate regime of strong bi-polarization emerges. This meta-stable state becomes persistent for large $\beta$ (see D).
	}
	\label{fig:FourTimeseries}
\end{figure*}

The model can give rise to a variety of collective phenomena.
In order to provide intuition about its dynamical behavior and to characterize the collective opinion processes that follow from different processing biases, we first look at a series of paradigmatic model realizations.
Figure \ref{fig:FourTimeseries} shows four individual realizations of the model with increasing $\beta$ from top to bottom.
The number of agents is set to $N = 1078$.
It shows (red shaded) how the distribution of opinions on the scale from -4 to 4 evolves due to repeated exchange of arguments.
Superimposed to this evolving distribution the mean opinion and the standard deviation (opinion diversity) is shown by the blue and red curves respectively.
Notice that the number of steps needed for convergence varies greatly across these four cases.
While 2000 iteration generally suffice to reach the stable state for moderate values of $\beta$ (panel B and C), the time period is extended to 40000 in the first and 20000 in the last example.
The plots are augmented with a characterization of different dynamical phases of the opinion process which is briefly described on the right hand side of the figure.

Panel A shows the behavior of the model in the absence of biased processing ($\beta = 0$).
In this scenario, repeated argument exchange gives rise to a process of diversity reduction (IIIa) by which all agents coordinate on the same arguments and hence opinions.
This process is very slow.
Almost 40000 iterations ($N/2$ pair interactions each) are needed to converge.
As the argument adoption is homogeneous ($p_{\beta} = 1/2$) independent of the attitude, this model falls into the class of consensus models for which convergence properties have been established in a seminal paper by \cite{DeGroot1974reaching}.
In particular, the probability of ending up with a specific opinion $o$ depends on the number of argument strings that are mapped to $o$, favoring moderate over extreme final opinions.

Panel B shows the effects of weak biased processing on the argument exchange process ($\beta = 0.4$).
In this scenario, the population quickly approaches one or the other extreme on the attitude scale with all agent strongly in favor or disfavor of the item at question.
For symmetric initial conditions the probability to end up in +4 or -4 is fifty-fifty.
Notice that compared to $\beta = 0$ convergence is extremely quick taking less than 500 iterations.
In the initial phase of the process (I) we observe a tendency of increasing opinion diversity due to biased processing.
This is followed by a phase of diversity reduction (IIIb) mimicking global choice shift towards one side.

If biased processing becomes larger (panel C and D) and crosses the critical value of $\beta^* = 1/2$ (see Section  \ref{sec:moderationpolarization}) a different dynamical phase emerges in the first period of the process.
Initially, the social pool of arguments is balanced and with strong processing bias agents are very likely to adopt arguments that support their initial attitudinal inclination and to reject arguments that challenge it.
As the analysis in Section \ref{sec:moderationpolarization} has revealed, there is a strong tendency of attitude polarization at the individual level.
That is, each single individual will strengthen its initial opinion and approach one or the other extreme (I).
Collectively, this leads to a state where approximately one half of the population approaches one extreme and the other half adopts an opposing view on the other side of the opinion spectrum (II).
We refer to this state as bi-polarization or collective polarization.
Once the system entered such a state, a very different process sets in which can be characterized as a competition for majority between the two opposing opinion groups.
As shown in panel D ($\beta = 1.2$), this phase of competition can be extremely persistent lasting more than 10000 iterations in this example.
At a certain point, however, due to rare random events by which individuals change side, one camp gains majority and the overall argument pool becomes biased into the respective direction.
Agents with the minority opinion are then more and more attracted due to this prevalence of majority arguments.

The phenomenological view that has been provided in this section aimed to convey basic intuition about the collective processes that emerge when biased processing is incorporated into argument communication models.
We have found that two remarkable transitions take place as the strength of the bias increases.
First, by the incorporation of a small processing bias, moderate consensus is no longer a stable outcome of the ACT models because the system quickly approaches a consensus at the extremes of the attitude scale. 
From the perspective of a group that faces a decision problem, weak biased processing hence enables a rather efficient group decision process.
Second, as biased processing increases, the system may enter a \revision{meta-stable} collective regime of bi-polarization with two groups of agents one strongly in favor and another one strongly against an issue (e.g. an energy technology).
\revision{This conflictual state becomes persistent as biased processing increases.}
Strong biased processing hence leads to a suboptimal group decision process \revision{as the group will need an extremely long time to arrive at a shared conclusion}.
We will provide a more detailed analysis of these two transitions in the following two sections.

\subsection{First transition: Weak biased processing leads to fast collective decisions}
\label{sec:firsttransition}

As shown in Fig. \ref{fig:FourTimeseries}, $\beta = 0$ (no bias) leads to a very long process in which the population is not clearly supporting or opposing an issue.
On the other hand, with $\beta = 0.4$, convergence times speed up by orders of magnitude leading to a very fast choice shift by all agents after which the group clearly favors one side over the other.
To better understand this transition, we run a computational experiment with focus on convergence times and the "sidedness" of the final group opinion varying the level of biased processing. 
In each simulation we measure the time that the system needs to converge to a stable state (consensus, phase IV in Fig. \ref{fig:FourTimeseries}) along with the absolute value of the respective consensual opinion.
Notice that the model is symmetric with respect to the attitude scale and converges to either side with equal probability.
The processing bias is varied from zero to $\beta = 0.4$ in steps of $1/60 \approx 0.0167$ (25 points) and at each sample point 100 simulations are performed.

\begin{figure}[ht]
	\centering
	\includegraphics[width=0.99\linewidth]{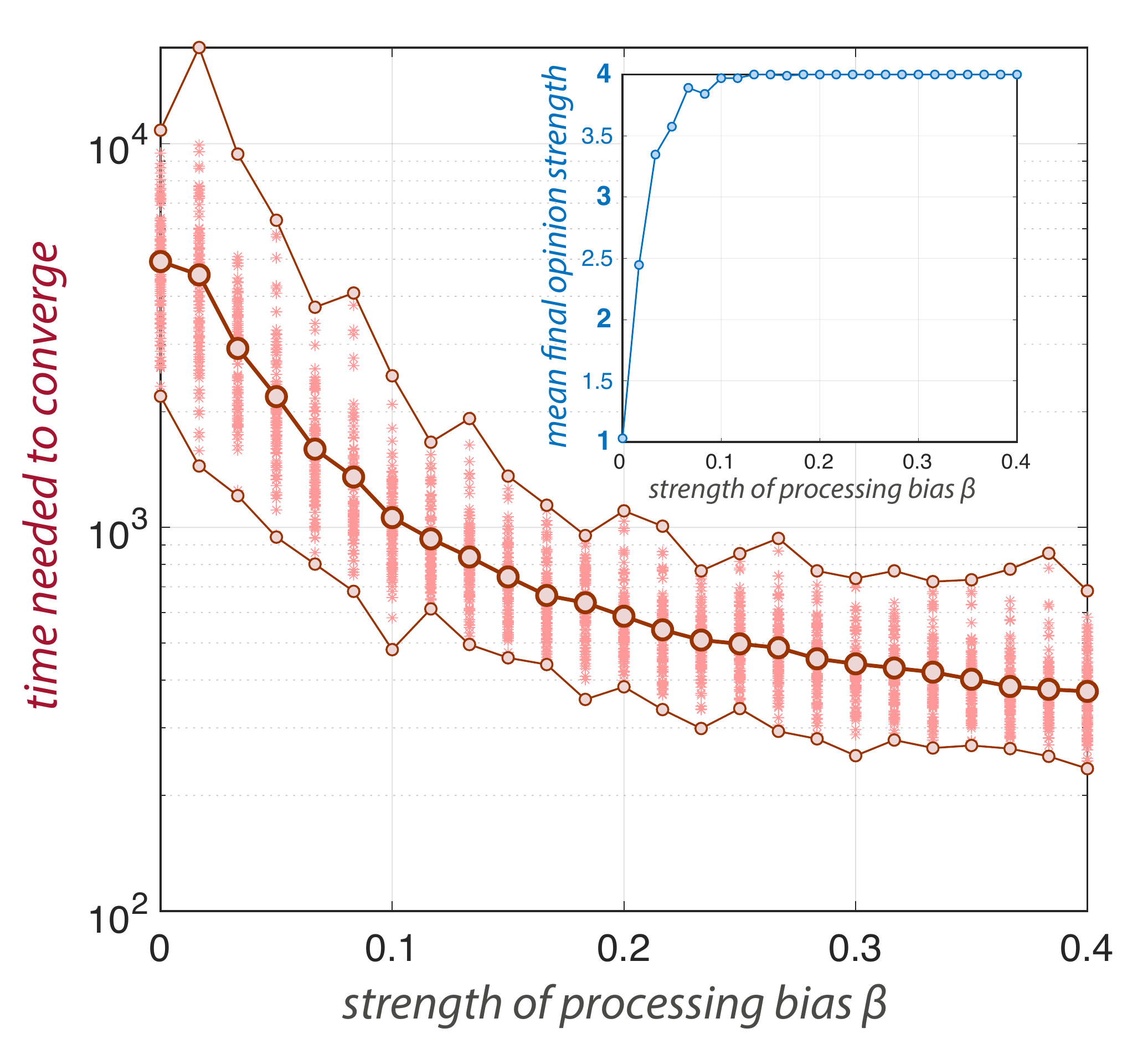}
	\caption{Time to reach a stable consensus profile as a function of $\beta \in [0,0.4]$. For each sample point a series of 100 simulation with $N = 100$ agents is performed. The mean value as well as the minimum and maximum values are shown along with the respective distribution of convergence times (light red). Inset: the inset shows the mean absolute value of the final group opinion. Under weak biased processing the final outcome shifts to the extremes of the opinion spectrum.
	}
	\label{fig:firsttransition}
\end{figure}

Fig. \ref{fig:firsttransition} shows the mean convergence time and the respective distribution over 100 runs on a logarithmic scale.
Minimal and maximal values are shown by the thin lines.
While it takes on average 5000 steps to convergence for $\beta = 0$ the mean number of iterations required to reach a stable profile is below 500 for $\beta \geq 0.25$.
Hence, weak biased processing significantly accelerates the consensus process.
In the inset of Fig. \ref{fig:firsttransition} the mean absolute value of the final group opinion is shown as a function of the processing bias.
For $\beta > 0.1$ the probability of ending up in a state different from -4 or +4 approaches zero, revealing a rather sharp transition towards an "extreme consensus".

We conclude that the inclusion of biased processing drastically affects the collective-level predictions of ACT models.
Even under very weak processing biases, moderate consensus is no longer a stable outcome of the model. 
Instead we observe quick convergence to one of the ends of the opinion spectrum where the entire group is strongly in favor or disfavor of the attitude object.
Hence, while groups without processing bias may remain in indecision for a long time not clearly favoring one side over the other, even small biases lead to a fast decision process with a clear outcome.
This has implications for previous theoretical work using ACT \citep{Maes2013short,Feliciani2020persuasion} and points towards an evolutionary function of biased processing at the group level.
We will discuss both points in the concluding part of the paper.

\subsection{Second transition: Strong biased processing leads to persistent intra-group conflict}
\label{sec:secondtransition}
The phenomenological analysis in Section \ref{sec:modelphenomenology} shows that biased processing alone may lead to persistent collective bi-polarization.
A particular composition of social groups formed around opinions may foster its emergence but it is not necessary.
To our knowledge, this is the first model that demonstrates this.
As biased processing increases, the collective behavior of the model undergoes a second transition from a regime of fast collective choice shift to a regime where enduring collective disagreement becomes likely.
Considering the initial periods (I) in Figure \ref{fig:FourTimeseries} suggests that the emergence of the disagreement regime rests on whether biased processing is strong enough to sustain attitude polarization at the individual level.
That is, we expect that collective opinion polarization becomes possible as $\beta$ crosses the critical values of $\beta^* = 1/2$ (cf. Figure \ref{fig:bifurcationplot}).
This section solidifies this result by a systematic computational experiment.

In order to systematically compare sets of model realizations regarding their potential to create collective polarization, we have to identify if a model trajectory has entered phase II in Figure \ref{fig:FourTimeseries}.
Many measures of opinion polarization have been conceived (see \cite{Bramson2016disambiguation} for an overview), and we define a conservative heuristic that captures the most important aspects.
We say that a system configuration is in phase II if the proportion of agents with extreme opinions on both sides of the opinion spectrum is larger than the proportion of agents with an opinion in between the two extremes.
To be precise, we split the opinion interval into three and count the number of agents with opinion -3 and -4 (negative extreme), the number of agents with opinion 3 and 4 (positive extreme), and the number of agents with an opinion from -2 to 2 (moderate).
If both the number of extremely positive agents \emph{and} the number of extremely negative agents exceed the number of moderates, we mark this configuration as bi-polarized.
Notice that this definition implies maximal opinion spread, high dispersion and low kurtosis to name a few measures used in the literature \citep{DiMaggio1996have,Bramson2016disambiguation}.

\begin{figure}[t]
	\centering
	\includegraphics[width=0.99\linewidth]{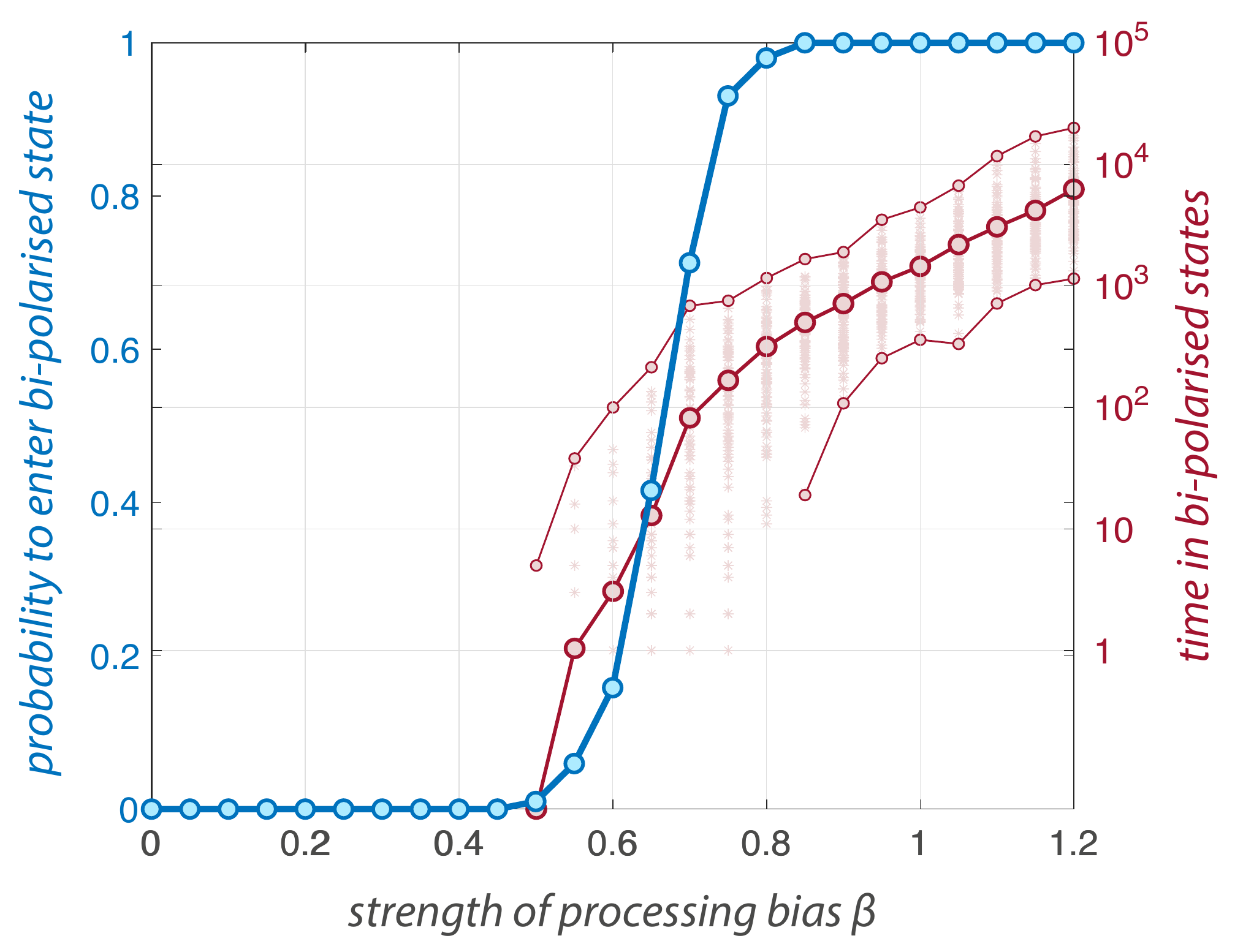}
	\caption{Probability that a system enters a state of collective polarization (blue) and the time it remains in such a state (red) as a function of biased processing $\beta$. Results based on 100 model runs per parameter with $N = 100$. For $\beta < 1/2$ the system does not polarize. The first (singular) instance of bi-polarization is observed at the critical value $\beta = 1/2$. As biased processing increases, the probability of a transient state of polarization increases and approaches 1 for $\beta > 0.8$. The persistence (number of time steps) of polarization is shown on a logarithmic scale. Mean persistence as well as the respective minimal and maximal number of steps are shown.
	}
	\label{fig:collectivepolarization}
\end{figure}

In the computational experiment we run a series of 100 realization of $N = 100$ agents for 100000 iterations and compute for each realization the number of time steps in phase II according the definition above.
The model parameter $\beta$ ranges from zero to 1.2 as before (Sections \ref{sec:microlevelimplications} and \ref{sec:calibration}) sampled with a step size of 0.05 (25 points).
For a given $\beta$, we assess (i.) the probability with which collective polarization emerges and (ii.) the number of time steps the system remains in this state (persistence).
Notice again that one iteration means $N/2$ interaction events in our implementation.
Both measures are shown in Figure \ref{fig:collectivepolarization}.

The blue curve shows the relative number of model runs which resulted in at least one temporal configuration that satisfies our polarization conditions.
The red curves show the respective number of time steps that a polarized state persisted for all 100 model runs highlighting the mean as well as the minimal and maximal value.
Notice the logarithmic scale on the right hand side of Figure \ref{fig:collectivepolarization}.
The regime $\beta < 1/2$ does not lead to transient states of stark disagreement.
The first instance is observed at a value of $\beta = 1/2$.
Hence, individual-level attitude polarization is necessary for collective polarization in our model.
However, the effect is not persistent and present only for 5 time steps in the respective model run.
In between $\beta = 1/2$ and $\beta = 0.8$ the probability of entering a state of bi-polarization becomes more and more likely and reaches one if $\beta > 0.8$.
At the same time, these transient opinion profiles of bi-polarization become truly persistent.
As shown by the distribution of time steps in polarization, persistence grows exponentially reaching several hundreds of iterations for $\beta = 0.8$ and several thousands of steps if $\beta > 1$ (compare also Figure \ref{fig:FourTimeseries}).
The analysis underlines that persistent collective polarization becomes likely if biased processing is strong.

It is worth noting at this point that a period of persistent bi-polarization is likely to (i.) transform the social organization of groups around opinion (homophily), may (ii.) lead to the emergence of symbolic leaders promoting group opinion (group identity) and (iii.) antagonistic relations across the groups (social polarization).
These processes are not integrated into the model, but they would all favor further persistence of collective polarization once such a pattern has emerged.
Our model shows that biased processing alone may be sufficient for the formation of camps that strongly support competing opinions.


\subsection{Influence of opinion homophily}

One of the most prevailing assumptions in opinion dynamics is that the interaction probability between two agents depends on the similarity of their opinions \citep{Axelrod1997dissemination,Hegselmann2002opinion,Deffuant2000mixing,Banisch2010acs}.
In previous ACT models \citep{Maes2013differentiation,Feliciani2020persuasion,Banisch2021argument} this homophily principle is considered the main mechanism responsible for collective bi-polarization.
As all previous ACT studies draw on homophily, it is important to understand the interplay of biased processing and homophily within this theoretical framework.

There are different ways to integrate opinion homophily into \revision{opinion dynamics models and ACT in particular.
First, opinion homophily may by implemented as the tendency to select similar interaction partners \citep[e.g.][]{Carley1991theory, Axelrod1997dissemination,Maes2013differentiation} or the strength of social influence \citep{Macy2003polarization,Flache2011small}.
Secondly, this similarity bias may be defined in absolute terms between pairs of opinions \citep{Deffuant2000mixing,Banisch2021argument} or relative to the entire population such that close partners are selected with a higher probability \citep{Carley1991theory, Maes2013differentiation}.
While the ACT model of \cite{Maes2013differentiation} proposes to operationalize homophily in relative terms as biased partner selection assuming that the opinions of all other agents are known,
}
\cite{Banisch2021argument} follow the tradition of bounded confidence models \citep{Hegselmann2002opinion,Deffuant2000mixing} and use a threshold on the opinion difference for a given pair of agents.
We adopt the latter approach here and assume that argument exchange takes place only if the opinion distance is below a certain threshold value.
\footnote{
\revision{
We have implemented the variant of \cite{Maes2013differentiation} into the model and simulations showed that the main results presented in this section are preserved. However, as this alternative implementation is significantly more costly in terms of computation, we decided to stick to the more simple bounded confidence variant.}
}

To analyze the impact of homophily in the refined ACT model a series of 100 simulations with $N = 100$ agents is performed for different values of $\beta$ ranging from $\beta = 0$ to $\beta = 0.8$.
As we are mainly interested in how far opinion homophily may foster the emergence of a bi-polarized group situation, we consider the fraction of simulation runs in which polarization (phase II in Fig. \ref{fig:FourTimeseries}, see previous section) can be observed in the transient dynamics.

\begin{figure}[t]
	\centering
	\includegraphics[width=0.99\linewidth]{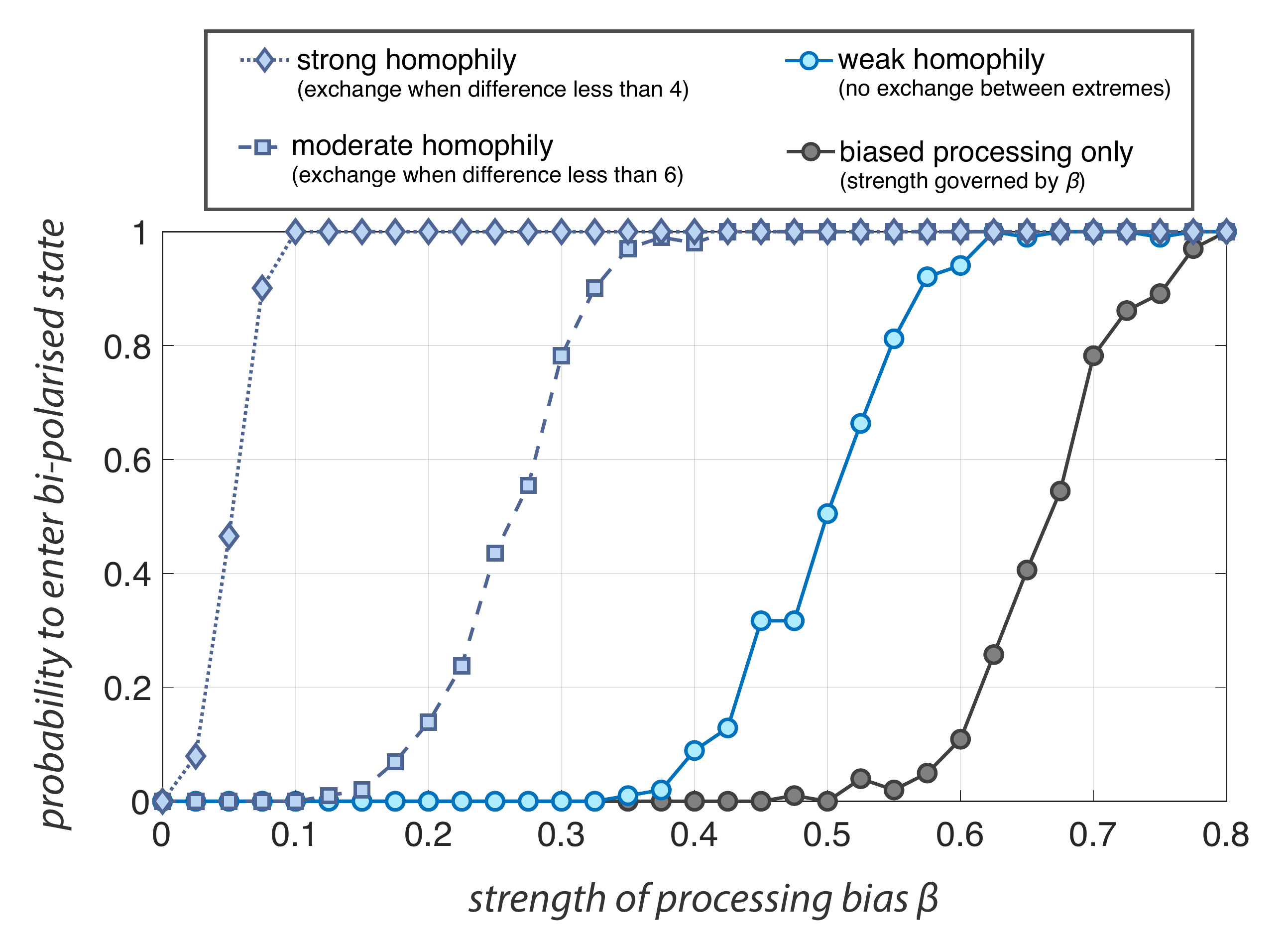}
	\caption{Probability that a system enters a state of collective polarization for different levels of biased processing and different levels of homophily. Results are based on 100 realizations with $N = 100$ agents. The figure compares the "base line" with biased processing only, and three model variants with increasing homophily.
	}
	\label{fig:homophily}
\end{figure}

In Fig. \ref{fig:homophily} the results are shown for three different values of the similarity threshold.
In our model attitude lie on a nine-point scale from -4 to +4. 
The weakest version of homophily is that agents strongly supportive of a given option (+4) do not enter into social exchange with agent strongly opposing it (-4), and vice versa.
That corresponds to a similarity threshold of $h = 8$.
Subsequently, we also show the results for $h=6$ and $h = 4$ as well as the "base line" without homophily (black).

The analysis shows that homophily makes polarization the most likely outcome of the collective process at significantly lower levels of biased processing.
Notably, group-level polarization can now emerge in the regime of individual-level attitude moderation $\beta < 1/2$.
Even in the rather weak form where exchange between the extremes is cut off a significant shift of the transition point to lower $\beta$ becomes visible.
Besides this shift, homophily also brings about a second qualitative change in collective model behavior: a completely bi-polarized opinion profile now becomes stable and phase II (Fig. \ref{fig:FourTimeseries}) enduring.
Once the system reaches a state of collective polarization with agents concentrated at both extremes of the opinion scale, homophily, as integrated here by a threshold function, makes further argument exchange impossible.
Hence, any profile in which agents are either completely in favor (+4) or against (-4) is an absorbing final state of the model dynamics. 

From the perspective of previous ACT models with $\beta = 0$, our results show that significantly lower homophily may result in a group split if a small amount of biased processing is introduced.
While in our model a rather restrictive threshold of $h = 4$ still leads to a moderate consensus, a small deviation in terms of attitude-dependent argument adoption makes bi-polarization the most likely outcome (dotted curve). 
This indicates that previous results are sensitive to small variations in assumptions about biased information processing.
As the empirical part of this paper demonstrates an increased micro-level validity of ACT when biased processing is included (at $\beta \approx 0.3$ for gas and biomass and of $\beta \approx 0.6$ for the remaining technologies), we have to assess whether previous conclusions drawing on ACT still hold in the presence of information processing biases.

\section{Concluding remarks}
\label{sec:discussion}

We conclude this paper by a summary and a brief discussion of its main contributions:

\begin{enumerate}
    \item 
    The paper presents a novel approach to combine an \revisionHS{empirical} experiment on argument persuasion with a computational theory of collective deliberation \revisionHS{to investigate the emergent phenomenon of opinion polarization processes}. It demonstrates that the theoretical framework of argument communication theory (ACT) can not only explain different dynamical phenomena in collective deliberation \citep{Maes2013differentiation,Maes2013short,Feliciani2020persuasion,Banisch2021argument}, but also provides a useful cognitive infrastructure to computationally map real experimental treatment. Starting from the theory, we develop a cognitively grounded statistical devise to assess the extent to which biased processing is involved in the experimentally observed attitude changes induced by conflicting but balanced arguments. We find that biased processing is relevant and improves the micro-level validity of argument-based models employed in the theory. With this coherent account bridging from experiments in Social Psychology to sociological models of collective opinion processes our work contributes to the major challenge of grounding social influence models rigorously in experimental data \citep[cf.][]{Flache2017models,Lorenz2021individual}, and proves ACT a useful candidate for achieving such an empirically more solid connection.
\item
    Following this program, we are able to clarify the relation between biased processing and attitude polarization at the individual level which has remained puzzling given the diverging empirical evidence through different persuasion experiments \citep[cf.][]{Corner2012uncertainty,Shamon2019changing}. Here we tackle this question from the point of view of computational agents employed in ACT and analyze how these cognitive agents would change opinions in a virtual experiment that matches closely to the real treatment. The theoretical response function for the expected attitude change  derived from that contains the strength of biased processing ($\beta$) as a free parameter. The theoretical analysis of this model shows that biased processing may lead to attitude moderation or attitude polarization if subjects are exposed to balanced arguments. Whether one or the other effect is observed depends crucially on the strength with which individuals engage in biased processing. In fact, our analysis reveals a sharp transition from moderation to polarization indicating that small, domain-specific variations in the strength of biased processing may result in qualitatively different patterns of attitude change, both consistent with our theory. Our work highlights that the question of whether biased processing leads to attitude polarization should not be asked in absolute but in relative terms and provides a theoretical explanation for why empirical evidence across different domains is mixed. 
\item
    Our empirical results concerning attitudes on electricity generating technologies show that the method advanced in this paper can provide a more refined, domain-specific understanding because it allows to measure the extent to which subjects engage in biased processing. On the entire data set, we find a clear signature of moderate biased processing at the margin of moderation and polarization. The independent analysis of the six groups that received arguments with respect to six different technologies reveals remarkable differences across topics. While the processing bias is in the regime of attitude moderation for gas and biomass, it is significantly higher and in the regime of polarization for coal, wind (onshore and offshore) as well as solar power. One possible explanation for this systematic differences is that beliefs on gas and biomass are less settled compared to the other four technologies and that beliefs regarding the latter are more strongly organized into coherent systems of beliefs \citep{Converse1964nature}.
\item
    The identification of the processing bias $\beta$ which matches best given experimental data for a specific attitude object is an efficient way to calibrate agent-based models of argument communication on the basis of balanced-argument experiments. The empirical analysis in the context of debates on different energy sources provides clear evidence that biased processing plays an important role in argument-induced attitude change and that its inclusion significantly improves the micro-level validity of the mechanisms assumed in current ACT models ($\beta = 0$). Given that different topics may elicit different degrees of biased processing (see previous point), the parameter $\beta$ provides a place for adjustment of a computational model with respect to opinions on a specific topic. Recent work has shown that ACT can incorporate arguments brought forward in real debates \citep{Willaert2021tracking}, and the experimental calibration regarding the argument exchange mechanism is a further step towards empirically-informed models of opinion dynamics. 
\item
    The analysis of the collective-level implications of our refined model shows that the incorporation of biased processing has tremendous effects on the predictions of ACT regarding the evolution of opinions within a group or a population. We observe two transitions. First, and somewhat surprisingly, weak biased processing accelerates group decision processes by orders of magnitude. While a group remains in a long period of indecision -- not clearly favoring one option over the other -- in previous models without bias, weak levels of biased processing quickly lead to a state in which all members jointly support one option. A second transition occurs if biased processing increases. Under strong biased processing the argument model leads to a persistent conflictual state of subgroup polarization.
\item 
    Our study hence shows that biased processing alone is sufficient for \revision{the emergence of} collective bi-polarization. While the  original model by \cite{Maes2013differentiation} has shown that polarization is possible under positive social influence if homophily is strong enough, our work shows that preferences for interaction with like-minded others are not necessary either. With that our work adds to the growing body of literature on mechanisms that contribute to societal polarization \citep[see ][ and references therein]{Flache2017models,Banisch2019opinion}. Moreover, while empirical plausibility of  inter-personal mechanisms of negative influence has been challenged \citep{Takacs2016discrepancy}, there is ample empirical evidence for the intra-personal mechanism of biased information processing that is at the core of our model. The experiment analyzed in this paper further provides convincing empirical ground for the microscopic validity of this mechanism.
\end{enumerate}

\revision{
In this paper, we concentrated on the effects of biased processing on individual attitude change and the resulting dynamics of collective opinion formation. There are many other social and cognitive mechanisms that are relevant for a better understanding of polarization dynamics which could be included into our model. As opinion homophily is a prevailing assumption in other models \citep{Axelrod1997dissemination,Hegselmann2002opinion,Flache2017models} and ACT in particular \citep{Maes2013differentiation,Feliciani2020persuasion,Banisch2021argument}, we have briefly addressed the interplay of biased processing and homophily, but refrained from incorporating further factors to keep the analysis clear and easy to interpret (see \cite{Lorenz2021individual} for recent work including quite a series of other factors). The incorporation of homophily has been based on a simple threshold model of "bounded confidence" \citep{Hegselmann2002opinion} and we showed even under the weakest threshold value cutting off interaction between the extremes the transient polarization pattern becomes stable. In this final part of the paper, we will discuss other factors that accelerate polarization in the setting of ACT.
}


In the context of the climate change debate, ample empirical evidence on biased information processing has been gathered in recent years. The experiment on which our analysis is relying \citep{Shamon2019changing} addresses the issue at the level of specific arguments providing a specific but at the same time systematic picture of how attitude extremity and direction impact biased processing. Another type of empirical evidence comes from a series of communication studies addressing the impact of selective media exposure in the climate change debate \citep{Feldman2011opinion, Hart2012boomerang,Nisbet2015partisan, Stroud2017understanding, Newman2018climate}. While it is long known that ideological affiliation is an important driver for media choice \citep{Lazarsfeld1944people}, a more refined picture of the interplay of attitudes and media choices has been obtained within the "reinforcing spirals framework" \citep{Slater2007reinforcing, Feldman2014mutual}. This theory posits a reinforcing feedback between selective media choice and biased information processing which over time increases informational fragmentation and opinion polarization. In future work, we will integrate selective exposure into our model to analyze the polarization potential of selective exposure in the presence of biased argument processing. Moreover, the reinforcing spirals model does not yet account for interpersonal influences \citep[p. 606]{Feldman2014mutual}. An operationalization within ACT overcomes this deficiency and provides a cognitive foundation that may proof useful to further disentangle the effects of biased processing, social influence and selective media exposure.

\revision{
While biased processing focuses on the perception and processing of information once individuals are exposed to a message, it seems reasonable to assume that people also tend to communicate arguments that are congruent with their opinion. Incorporating "biased argumentation" into our model is rather straightforward and could be based on the procedure that now governs argument adoption (Eq. \ref{eq:argadoptionmodel2}). That is, the probability to communicate congruent arguments is biased with a certain strength, say $\gamma$. This would have strong implications in the ACT setting of interpersonal communication. Most notably, it would shift the critical value at which collective polarization emerges to significantly lower levels, because the effective bias increases by $\beta + \gamma$. However, argument production is more difficult to address in randomized experiments 
and a highly standardized balanced argument setting needed to assess an argument production bias will require a very careful design.


A promising direction for future research is the incorporation of more cognitive complexity into the model. Issues and arguments are not independent from one another: certain claims may support or attack other arguments to form complex systems of beliefs \citep{Converse1964nature,Dalege2016toward,Boutyline2021cultural,Taillandier2021introducing}. The cognitive agent model used in this paper has been derived from the principles of cognitive coherence affecting the evaluative part $V(a)$ of the argument adoption probability (Eq. \ref{eq:argadoptionmodel2}). In a setting where arguments are interrelated, biases are therefore amplified favoring argument configurations of high internal consistency. This might be a key to better understand the cognitive power of conspiracy narratives. In the context of our application, it also provides one possible explanation for why biases are different for different issues (Section \ref{sec:issuedifferences}). Namely, one reason for these differences may be that beliefs and arguments on coal, wind and solar-based technologies -- with more tendency of biased processing -- form more densely connected networks of compatible and incompatible beliefs favoring a stronger evaluation bias. The fact that debates around the two technologies that reveal less tendency of biased processing -- gas and biomass -- have come to public attention only more recently compared to coal, wind and solar power further supports this hypothesis.

A series of further interesting questions for model analyzes relates to the incorporation of more agent heterogeneity, the actual tenet of agent-based modeling. First of all, our model assumes that social interaction is completely random. While random mixing might be a plausible assumption in small group discussions, it is no longer plausible for larger populations where social networks typically exhibit considerable degree heterogeneity, local clustering and community structure \citep{Wasserman1994social,Newman2002random,Borgatti2009network}. 
On modular networks with weak ties across cohesive communities we would observe bi-polarization within the subgroups if the bias $\beta$ is high. If biased processing is moderate or low, we observe fast convergence of community opinions approaching one or the other extreme with equal chance. Hence, in this case the model behaves similar to social feedback models reinforcing opinions within cohesive groups \citep{Banisch2019opinion}.
Furthermore, it seems reasonable that actors interpret arguments not only in terms of what they already believe, but also in terms of who is offering the argument and how trustworthy the information provided by a sender is for the receiver. These „source effects“ have received considerable attention in psychological research on persuasion \citep{Wilson1993source}. In social influence network theory \citep{Friedkin2011social}, asymmetric status characteristics (expertise, authority, and other forms of power) are subsumed into the influence network \citep[p. 861/62]{Friedkin1999choice}. In a similar way, heterogeneous and possibly asymmetric social relations could be integrated into our model.


A second type of heterogeneity that deserves further analysis is heterogeneity with respect to the strength of biased processing and the underlying networks of cognitive-affective associations. We have tested the effect of drawing individual $\beta$'s from a normal distribution and found the results shown in Section \ref{sec:secondtransition} reproduced. Convergence times, meta-stability (persistence) of bi-polar configurations, as well as probability to enter such a state are not significantly affected. But other patterns of heterogeneity potentially resulting from differences in the argumentative associations different individuals have internalized should be addressed. In particular, the notion of "stubborn" \citep{Acemoglu2013opinion}, "extremist" \citep{Deffuant2002how} or "zealot" \citep{Mobilia2003does} agents with a fixed extreme opinion can be refined and relates to agents with a stronger feel of challenge when confronted to counter arguments. On the basis of preliminary simulations it seems particularly interesting to study the effect of those agents in the early phase of the group deliberation process.
}

\revision{Finally}, this work inspires new thought about potential evolutionary origins of biased information processing. Groups often face situations in which cohesive action is needed and where choosing any out of a set of alternatives is better than taking no action at all. We found that a certain level of biased processing is very efficient from the group perspective in this specific sense. For a value close to the critical $\beta = 1/2$, the model predicts a very quick process in which one alternative becomes jointly supported by the entire group. Weaker biases slow down the group decision process and the group may remain undecided for a long time. Stronger biases, on the other hand, may lead to polarization and conflict. This points towards an evolutionary function of biased processing and selective information processing more generally: a specific level of bias may have evolved due to the selective pressures on a group’s ability to cohesively take joint action. 

All in all, this paper shows that biased processing increases the micro--validity of ACT and has a strong impact on its macro--level predictions. Future work has to clarify whether previous conclusions drawing on the theory still hold after our empirical refinement.

\subsection*{Acknowledgements}

Thanks to Stefan Westermann for pointing at the bifurcation analysis in Section \ref{sec:microlevelimplications}.
We also acknowledge helpful comments by three anonymous referees.
This project has received funding from the European Union’s Horizon 2020 research and innovation programme under grant agreement No 732942 (Odycceus -- Opinion Dynamics and Cultural Conflict in European Spaces).


\end{document}